\documentclass[prd, preprintnumbers, showpacs, nofootinbib, superscriptaddress,notitlepage, 11pt]{revtex4-1}

\usepackage[colorlinks=true,linkcolor=blue,urlcolor=magenta,citecolor=blue]{hyperref}
\usepackage{ulem}

\usepackage[english]{babel}
\usepackage[utf8]{inputenc}
\usepackage{mathtools}
\usepackage{physics}
\usepackage{xcolor}
\usepackage{multirow}
\usepackage[a4paper, left=2.5cm, right=2.5cm, top=2.5cm, bottom=2.5cm]{geometry}
\usepackage{adjustbox}
\usepackage{placeins}
\usepackage[T1]{fontenc}
\usepackage{lipsum}
\usepackage{csquotes}
\usepackage{amsmath,amssymb,amsthm}  
\usepackage{amsfonts}
\usepackage{mathrsfs}  
\usepackage{latexsym,bm}
\usepackage{graphicx}   
\usepackage{float}    
\usepackage{wrapfig}    
\usepackage{indentfirst}
\usepackage{slashed}  
\usepackage{units} 
\usepackage{tabularx} 
\usepackage{fancybox} 
\usepackage{array}    
\usepackage{tabularx}    
\usepackage{booktabs}    
\usepackage{subcaption}
\usepackage{hyperref}
\usepackage{caption}
\usepackage{makecell}
\newcommand{\beq}{\begin{eqnarray}}
\newcommand{\eeq}{\end{eqnarray}}
\newcommand{\ben}{\begin{enumerate}}
\newcommand{\een}{\end{enumerate}}
\newcommand{\non}{\nonumber\\ }

\newcommand{\xsl}{ x \hspace{-2.2truemm}/ }

\newcommand{\etasl}{ \eta \hspace{-2.2truemm}/ }
\newcommand{\psl}{ p \hspace{-2.2truemm}/ }

\allowdisplaybreaks[4]

\begin{document}

\begin{titlepage}
\title{\LARGE Semileptonic Decays of $D \to \rho l^+ \nu$ and $D_{(s)} \to K^\ast l^+ \nu$ from Light-Cone Sum Rules}

\author{Wang Lin}
\affiliation{School of Physics and Electronics, Hunan University, 410082 Changsha, China}

\author{Xiao-En Huang}
\affiliation{School of Physics and Electronics, Hunan University, 410082 Changsha, China}

\author{Shan Cheng}\email[Corresponding author: ]{scheng@hnu.edu.cn}
\affiliation{School of Physics and Electronics, Hunan University, 410082 Changsha, China}
\affiliation{School for Theoretical Physics, Hunan University, 410082 Changsha, China}
\affiliation{Hunan Provincial Key Laboratory of High-Energy Scale Physics and Applications, 410082 Changsha, China}

\author{De-Liang Yao}\email[Corresponding author: ]{yaodeliang@hnu.edu.cn}
\affiliation{School of Physics and Electronics, Hunan University, 410082 Changsha, China}
\affiliation{School for Theoretical Physics, Hunan University, 410082 Changsha, China}
\affiliation{Hunan Provincial Key Laboratory of High-Energy Scale Physics and Applications, 410082 Changsha, China}

\date{\today}

\begin{abstract}

We investigate the semileptonic decays of charmed mesons to light vector mesons within the framework of light-cone sum rules. Our calculation is performed at leading order in QCD coupling, incorporating contributions up to twist-five accuracy from both two-particle and three-particle light-cone distribution amplitudes. Transition form factors are predicted twist by twist to assess the convergence property of the operator product expansion. It is verified that the twist-four and twist-five contributions are indeed negligible for all the decays under consideration. Twist-three dominance is observed for some of the form factors, subject to heavy quark effective field theory interpretation. Branching ratios for the decays $D^+ \to \rho^0 \ell^+\nu_\ell$ , $D_s^+ \to K^{\ast0} \ell^+\nu_\ell$, $D^0 \to K^{\ast-} \ell^+\nu_\ell$ and $D^+ \to \bar{K}^{\ast0} \ell^+\nu_\ell$ are obtained, and a $10\%$--$20\%$ discrepancy  from experimental measurements is found. Our finding indicates that the resonant-width and non-resonant QCD backgrounds effects should be potentially significant, implying the necessity to further implement their contributions in future precision studies of the semileptonic charm decays.

\end{abstract}
\maketitle

\end{titlepage}

\section{Introduction}

Semileptonic decays of charmed mesons serve as powerful probes for the exploration of physics beyond the Standard Model (SM). On the one hand, they provide precise determinations of the Cabibbo-Kobayashi-Maskawa (CKM) matrix elements $\vert V_{cd} \vert$ and $\vert V_{cs} \vert$~\cite{HFLAV:2022esi,ParticleDataGroup:2024cfk}, which in turn can be used to test the unitarity of the CKM matrix. On the other hand, they also enable stringent tests of lepton flavor universality (LFU) through observables such as the ratio ${\cal R}_{\mu/e}^{c \to d}$~\cite{Isgur:1988gb,BESIII:2018nzb,Ablikim:2020tmg}. Precise calculations of the heavy-to-light form factors (FFs), i.e. functions of the momentum transfer squared $q^2$, are essential to describe these decays, as they encode the underlying QCD dynamics of the hadronic transition. The short-distance quark-gluon dynamics is facilitated by the hard scales, namely, the masses of the $W$ boson mediating the weak interaction and the charm quark, which guarantee the perturbative treatment of the form factors in QCD for a certain kinematical region of $q^2$. 

Compared to the semileptonic decays of $D \to P l^+ \nu$, the $D \to V l^+ \nu$ transitions offer multidimensional observables for experimental and theoretical investigations, where $P$ and $V$ denote pseudoscalar and vector mesons, respectively. This is due to the fact that the vector meson is usually unstable and subsequently decays into two pseudoscalar mesons ($P_1P_2$ with the momentum $k=k_1+k_2$). More specifically, one has $D \to V l^+ \nu \to P_1P_2l^+ \nu$, which is called $D_{\ell 4}$ decay for short. For instance, the process of $D \to \rho l \nu$ is experimentally observed as a four-body semileptonic decay of the type $D \to \pi\pi l \nu$~\cite{Faller:2013dwa,Kang:2013jaa}, where the final $\pi\pi$ system is in $P$ wave. It should be emphasized that
the FFs involved in the $D_{\ell4}$ decays depend not only on the momentum transfer squared $q^2$ but also on the $P_1P_2$ invariant mass squared $k^2$. Furthermore, the $D_{l4}$ decays provide access to multiple angular observables \cite{BESIII:2018qmf,BESIII:2024lxg}, such as the forward-backward asymmetry and other differential distributions, enabling more sensitive tests of the SM. 

In recent years, theoretical efforts~\cite{Cheng:2017smj,Hambrock:2015aor,Cheng:2017sfk,Cheng:2019hpq,Boer:2016iez,Leskovec:2025gsw,Herren:2025cwv} have been extensively devoted to the study of four-body leptonic decays of $B$ mesons ($B_{l4}$ decays). 
Amongst them, special attention has been paid to the decay mode $B \to \pi\pi l\nu$, due to its importance for the extraction of the modulus of $V_{ub}$ CKM matrix element; see e.g., Ref.~\cite{Cheng:2025hxe} for more discussions. However, in the charm sector, the $D_{\ell4}$ decays remain less investigated and the relevant $D \to P_1P_2$ form factors are poorly known. The first attempt is made for $D\to \pi\pi \ell\bar{\nu}$ in Ref.~\cite{Shi:2017pgh} by imposing unitarized chiral perturbation theory (ChPT). As expected, therein, the obtained FFs are valid only for large $q^2$ and hence cannot be applied to describe experimental data, e.g., by the BESIII collaboration~\cite{BESIII:2018qmf,BESIII:2024lxg,BESIII:2024qnx}, of the differential decay rates. To achieve a good description of experimental data in the whole kinematical region in the future, light-cone-sum-rule (LCSR) determinations of the $D \to K \pi$ and $D \to \pi\pi$ form factors at low $q^2$ are indispensable~\cite{D2pipi-LCSRs}. 

Nevertheless, thanks to the vector meson dominance ansatz, the $D_{\ell 4}$ decays are expected to be dominated by the $P$-wave of the final meson pair, where the vector meson $V$ behaves as an intermediate state. Especially when the vector meson under consideration is a standard Breit-Wigner resonance with narrow width, the transition amplitude of $D_{\ell 4}$ can be factorized as a product of the weak transition amplitude of $\mathcal{A}_{D\to V\ell \nu}$, the Breit-Wigner propagator ${\rm BR}_V$ and the strong decay amplitude $\mathcal{A}_{V\to P_1 P_2}$. Since the latter two terms are well constrained by experimental data, the key challenge lies in calculating the $D \to V\ell \nu$ sub-amplitude, especially the $D\to V$ transition FFs. 
These FFs have been studied through various theoretical frameworks, such as QCD-based approaches~\cite{Ball:1993tp,Wu:2006rd,Feldmann:2017izn,Fu:2020vqd}, quark-model formalisms~\cite{Scora:1995ty,Cheng:2003sm,Verma:2011yw,Cheng:2017pcq,Soni:2018adu,Ivanov:2019nqd,Faustov:2019mqr}, and effective field theories~\cite{Fajfer:2005ug,Sekihara:2015iha}.
In the current work, we prefer the QCD LCSR method, 
which is a rigorous framework in the sense that it faithfully respects the first principles of QCD. 

By employing the operator product expansion (OPE) near the light-cone, the LCSRs approach admits a systematic organization of various twist contributions and a direct relation to the underlying QCD parameters of hadrons. Furthermore, the theoretical uncertainties can be well controlled given that a proper approximation scheme is used. 
The application of LCSRs to the $D \to V$ transitions, however, is less explored compared to the $D_{(s)} \to P$ counterparts with $P$ being pseudoscalars like $\pi,K,\eta^{(')}$ \cite{Khodjamirian:2000ds,Khodjamirian:2009ys,Offen:2013nma,Yao:2018tqn}. 
This relative scarcity can be attributed to the larger masses and the widths of the vector mesons in comparison to the pseudoscalar mesons, which make the calculation of $D \to V$ FFs much more challenging. Besides, it is also well acknowledged that the convergency of the series expansion in twist for the $D \to V$ FFs is not as good as that in the $D \to P$ case. Yet, for all that, the utility of LCSR in the study of $D\to V$ decays is being powerful. Reliable predictions of $D\to V$ FFs have been achieved in literature~\cite{Wu:2006rd,Fu:2020vqd}, albeit at twist-4 accuracy. In this work, we are going to conduct a systematical LCSR analysis of the $D\to V$ FFs, with $D\in\{D,D_s\}$ and $V\in\{\rho, K^\ast\}$, up to the order of twist five. 
Our investigation takes into account contributions both from the two-particle and three-particle light-cone distribution amplitudes (LCDAs). 
It is found that the two-particle twist-5 corrections contribute slightly, merely a few percent. 
In contrast, the three-particle contributions turn out to be substantial, comparable in magnitude to those from two-particle twist-3 LCDAs.

This manuscript is structured as follows. 
In Section \ref{ffs-LCSRs}, we revisit the derivation of $D \to V$ transition FFs within the LCSR framework, using LCDAs of the vector mesons as inputs. 
Section~\ref{ffs-numerics} comprises numerical results and discussion of the FFs, 
therein the subsection \ref{semileptonicdecays} presents our predictions for the semileptonic decays $D \to (\rho, K^\ast) l \nu$ and $D_s \to K^\ast l\nu$. 
We summary in section \ref{sec:summary}. For completeness, the explicit expressions of vector-meson LCDAs are given in Appendix~\ref{app:LCDAs}.

\section{Derivation of $D \to V$ form factors from LCSR}\label{ffs-LCSRs}
\subsection{Definition of form factors}
The $D \to V$ transition FFs are defined through the matrix element of the $c\to q^\prime$ weak current sandwiched between the initial $D$ meson and the final vector meson states,
\beq \langle V(p, \eta^\ast) \vert \Gamma^\mu \vert D(p_D) \rangle = &-\left( m_D + m_V \right) \mathcal{O}_{A_1}^\mu  A_1(q^2) +  \dfrac{\mathcal{O}_{A_2}^\mu}{m_D + m_V} A_2(q^2) \non
&+ \dfrac{2m_V}{q^2}\mathcal{O}_{A_3}^\mu \left( A_3(q^2) - A_0(q^2) \right) - \dfrac{\mathcal{O}^\mu_V}{m_D +m_V} V(q^2). \label{eq:FFs-def1} \eeq
where $\Gamma^\mu = {\bar q}^\prime \gamma^\mu \left( 1 - \gamma_5 \right) c$ with $q^\prime\in\{d,s\}$, and the Lorentz operators are given by
\begin{align}
    \mathcal{O}_{A_1}^\mu = i \eta^{\ast \mu}
    \ ,\quad  \mathcal{O}_{A_2}^\mu =i \left( p_D +p \right)^\mu\eta^\ast \cdot q\ ,\quad
    \mathcal{O}_{A_3}^\mu =i q^\mu (\eta^\ast \cdot q)
    \ ,\quad
   \mathcal{O}^\mu_V= -2 \varepsilon^{\mu\nu\rho\sigma} \eta^\ast_\nu q_{\rho} p_\sigma \ .\label{eq.Lor.operators}
\end{align}
Here, $p_D$ and $p$ are four momenta of the charmed $D$ meson and vector meson\footnote{In this work, we focus exclusively on the vector mesons $V = \rho, K^\ast$, with this designation implied hereafter unless otherwise specified.}, respectively. The momentum transferred by the weak current is $q = p_D - p$. The polarization vector of the vector meson is denoted by $\eta^\ast$, satisfying the orthogonality condition $\eta^\ast \cdot p = 0$. There exists a kinematical constraint, which reads
\begin{align}
    A_3(q^2) =\frac{1}{2m_V} \left[(m_D + m_V) A_1(q^2) - (m_D - m_V) A_2(q^2) \right]\ .
\end{align}
Therefore, only four of the FFs defined in Eq. (\ref{eq:FFs-def1}) are independent, and are usually chosen to be $\{V, A_0, A_1, A_3 \}$. The befinit of the FFs defined in Eq. (\ref{eq:FFs-def1}) is that they can be straightforwardly related to the helicity amplitudes (c.f. Eq.~\eqref{eq.hel.amp}), offering a frame-independent description of the decay dynamics.

Alternatively, the transition matrix element can be decomposed in an orthogonal basis of Lorentz operators as
\beq \langle V(p, \eta^\ast) \vert \Gamma^\mu \vert D(p_D) \rangle = p_1^\mu {\cal V}_1(q^2) + p_2^\mu {\cal V}_2(q^2) + p_3^\mu {\cal V}_3(q^2) + p_P^\mu {\cal V}_P(q^2), \label{eq:FFs-def2} \eeq
where the four independent momenta are defined by
\beq && p_1^\mu = 2 \varepsilon^{\mu \nu \rho \sigma} \eta^\ast_\nu p_\rho q_\sigma, \qquad  
p_2^\mu = i \left\{ \left(m_D^2 - m_V^2 \right) \eta^{\ast \mu} - (\eta^\ast \cdot q)\left(p+p_D \right)^\mu \right\}, \non
&& p_3^\mu = i \left( \eta^\ast \cdot q \right) \left\{ q^\mu - \frac{q^2}{m_D^2 - m_V^2} \left( p + p_D \right)^\mu \right\}, \qquad 
p_P^\mu = i (\eta^\ast \cdot q) q^\mu. \label{eq:orth-momenta} \eeq 
The relations between the FFs, $\{V, A_0, A_1, A_3 \}$ in the helicity-related basis, and the ones, $\{{\cal V}_1, {\cal V}_2, {\cal V}_3, {\cal V}_P \}$ in the orthogonal basis, are given by
\beq &&{\cal V}_P(q^2) = - \frac{2 m_V}{q^2} A_0(q^2)\ , \qquad {\cal V}_1(q^2) = - \frac{V(q^2)}{m_D + m_V}\ , \non
&&{\cal V}_2(q^2) = - \frac{A_1(q^2)}{m_D - m_V}\ , \qquad 
{\cal V}_3(q^2) = \frac{2m_V}{q^2} A_3(q^2)\ . \label{eq:relation-FFs} \eeq

\subsection{Form factor densities}
The LCSR calculation of $D \to V$ transition FFs starts from the correlation function 
\beq C[\Gamma^\mu] = i \int d^4x e^{-i q \cdot x} \langle V(p,\eta^\ast) \vert T \{ \Gamma^\mu(x), \Gamma_5(0) \} \vert 0 \rangle, \label{eq:CF} \eeq 
where the $c \to d/s$ weak current $\Gamma^\mu$ is correlated with a pseudoscalar current $\Gamma_5=m_c {\bar c} i \gamma_5 q_2$, which has the same quantum numbers as the $D$ or $D_s$ meson. 
In the kinematical region where $q^2 \ll m_c^2$ and $(p+q)^2 \ll m_c^2$, the $c$ quark, emitted from the pseudoscalar current and subsequently absorbed by the weak current, becomes highly virtual. Consequently, it propagates at small average $x^2$ and the corresponding propagator can be written as 
\beq s_c(x,0,m_c) \equiv -i \langle 0 \vert T \{ c_i(x), {\bar c}_j(0) \} \vert 0 \rangle= \frac{-i m_c^2}{4\pi^2} \left[ \frac{K_1(m_c\sqrt{\vert x^2 \vert})}{\sqrt{\vert x^2 \vert}} + \frac{i \xsl K_2(m_c \sqrt{\vert x^2 \vert})}{\vert x^2 \vert} \right] \delta_{ij} \non
- \frac{1}{16\pi^2} \int_0^1 dv \left[ m_c K_0(m_c\sqrt{\vert x^2 \vert}) (\mathbf{G} \cdot \sigma) 
+ \frac{im_c K_1(m_c\sqrt{\vert x^2 \vert})}{\sqrt{\vert x^2 \vert}} \left( {\bar v} \xsl (\mathbf{G} \cdot \sigma) + v (\mathbf{G} \cdot \sigma) \xsl \right) \right] \delta_{ij}. \eeq
where $\mathbf{G}^{\mu\nu}(vx)=\mathbf{G}^{\mu\nu a}(vx){\lambda^a}/{2}$ and $\sigma^{\mu\nu}=\frac{i}{2}[\gamma^\mu,\gamma^\nu]$.
Here $K_i$ denotes the modified Bessel function of the second kind. The first and second terms correspond to the free-charm quark propagator and the quark-gluon interaction at leading power of the strong coupling $g_s$, respectively. 
Meanwhile, the light quark-antiquark pair forming the vector meson is emitted with near-lightlike separation, $x^2 \sim 0$. Hence, the vector meson dynamics is thoroughly characterized by its LCDAs, which are collected in Appendix~\ref{app:LCDAs}. In the above-mentioned kinematical regime, the OPE near the light-cone can be effectively applied to the correlation function. Consequently, the valid domain of OPE for $D_{(s)} \to V l\nu$ decays corresponds to $0 \leq q^2 \leq m_c^2 - 2 m_c \chi$, where $\chi \sim 500$ MeV is a typical hadronic scale parameter.

With the help of OPE at the quark-gluon level, the correlation function can be cast as a twist-ordered convolution of hard coefficients with LCDAs, 
\beq C[\Gamma_\mu] &=&
\sum_{{t}=2} \int_0^1 du \, T_{\mu}^{(t)}(u,q^2,(p+q)^2) \, \phi^{ (t)}_{2p}(u) \non &+& \sum_{{t}=3}\int_0^1 du \int_0^u {\cal D} \alpha_i \, T_{\mu}^{(t)}(u, \alpha_i, q^2, (p+q)^2) \, \phi_{3p}^{(t)}(\alpha_i), \label{eq:CF-OPE} \eeq
where the sums are taken over twists. Here the generic symbols, $\phi_{2p}^{(t)}$ and $\phi_{3p}^{(t)}$, stand for two-particle (${\bar q}q$) and three-particle (${\bar q}qg$) LCDAs of twist-$t$, respectively. Therefore, in the above equation, the frist and second terms correspond to the contributions from two-particle and three-particle LCDA configurations, respectively. Note that $T_\mu^{(t)}$ are known functions from LCSR, which can be further decomposed according to their pertinent Lorentz structure. 

Transparently, the correlation function has the same Lorentz structure as the transition matrix element in~Eq.~\eqref{eq:FFs-def1}. Namely, one has
\beq C[\Gamma^\mu] = C_1 \mathcal{O}_V^\mu + C_2 \mathcal{O}_{A_1}^\mu +  C_3 \mathcal{O}_{A_3}^\mu  +  C_4 (\mathcal{O}_{A_2}^\mu -\mathcal{O}_{A_3}^\mu )\ , 
 \eeq 
where $C_{1,2,3,4}$ are called FF densities and the operators are specified in Eq.~\eqref{eq.Lor.operators}. The densities $C_{i=1,2,3,4}$ are obtained directly from the OPE calculations, whose explict expressions read 
\beq &&C_1 = \int_0^1 du \Big[ - \frac{f_V^\perp m_c}{2} \frac{\phi_2^\perp(u)}{\triangle} 
+ \frac{m_V^2 f_V^\perp m_c}{8} \left( \frac{1}{\triangle^2} +\frac{2m_c^2 }{\triangle^3} \right) \phi_4^\perp(u) \non
&& \hspace{1.2cm} - \frac{m_c^2 f_V m_V}{4} \frac{\tilde{\psi}_3^\perp(u)}{\triangle^2} 
+ \frac{3 m_c^4 f_V m_V^3}{8} \frac{\tilde{\psi}_5^\perp(u)}{\triangle^4} \Big] \non
&& \hspace{0.6cm} - f_V^\perp m_V^2 m_c \int {\cal D}\alpha_i \Big[ \frac{S(\alpha_i) - (1-2v) \tilde{S}(\alpha_i)}{2 \square^2}  + \frac{T_1^{(4)}(\alpha_i) - T_2^{(4)}(\alpha_i) + T_3^{(4)}(\alpha_i) - T_4^{(4)}(\alpha_i) }{\square^2} \non
&& \hspace{1.2cm} - \left( 1 + \frac{m_c^2 -q^2 + \left( \alpha_1 + v \alpha_3 \right)^2 m_V^2}{\square} \right)
\frac{4v \left(T_3^{(4)}({\bar \alpha}_1, \alpha_3) - T_4^{(4)}({\bar \alpha}_1, \alpha_3) \right) }{\left( \alpha_1 + v \alpha_3 \right) \square^2} \Big]\ , \nonumber\\
&&C_2 = \int_0^1 du \Big\{ \frac{m_c^2 f_V m_V}{m_D^2 -m_V^2} \Big[ - \frac{\phi_3^\perp(u)}{\triangle} 
+ \frac{m_c^2 m_V^2}{2} \frac{\phi_5^\perp(u)}{\triangle^3} + \frac{m_V^2}{\triangle^2} \, I_2[\psi_4^\perp - 2 \phi_3^\perp + \phi_2^\parallel](u) \Big] \non 
&& \hspace{1.2cm} + \frac{f_V^\perp m_c}{m_D^2 -m_V^2} \Big[ \left( 1 - \frac{m_c^2 - q^2 + u^2m_V^2}{\triangle} \right) \frac{\phi_2^\perp(u)} {2u} \non
&& \hspace{1.2cm} - \left( \frac{1}{\triangle} + \frac{m_c^2 + q^2 - u^2m_V^2}{\triangle^2} - \frac{2m_c^2 \left( m_c^2 -q^2 + u^2m_V^2 \right)}{\triangle^3} \right) 
\frac{m_V^2 \phi_4^\perp(u)}{8u} \non
&& \hspace{1.2cm} - \left( \frac{1}{\triangle} - \frac{m_c^2 - q^2 + u^2m_V^2}{\triangle^2} \right) 
\frac{m_V^2}{u} \, I_2[\phi_3^\parallel - \frac{\phi_2^\perp}{2} - \frac{\phi_4^\perp}{2}](u) \non
&& \hspace{1.2cm} + \left( \frac{1}{\triangle} + \frac{2m_c^2}{\triangle^2} \right) \frac{m_V^2 }{2} \, I_1[\psi_4^\perp - \phi_2^\perp](u) 
- \frac{m_V^2}{2} \frac{\tilde{\psi}^\parallel_3(u)}{\triangle} \Big] \Big\} \non
&& \hspace{0.6cm} - f_V^\perp m_V^2 m_c \int {\cal D}\alpha_i 
\left( 1 + \frac{m_c^2 -q^2 + \left( \alpha_1 + v \alpha_3 \right)^2 m_V^2}{\square} \right)
\Big[ \frac{\left((1-2v) S(\alpha_1) - \tilde{S}(\alpha_i) \right) }{2 \left( \alpha_1 + v \alpha_3 \right) \square} \non
&& \hspace{1.2cm} + \frac{T_1^{(4)}(\alpha_i) - (1-6v) T_2^{(4)}(\alpha_i) + 
T_3^{(4)}(\alpha_i) - (1-2v) T_4^{(4)}(\alpha_i) }{\left( \alpha_1 + v \alpha_3 \right) \square} \non
&& \hspace{1.2cm} - \left( 1 + \frac{m_c^2 -q^2 + \left( \alpha_1 + v \alpha_3 \right)^2 m_V^2}{\square} \right) 
\frac{4v \left(T_3^{(4)}({\bar \alpha}_1, \alpha_3) + T_4^{(4)}({\bar \alpha}_1, \alpha_3) \right)}{ \left( \alpha_1 + v \alpha_3 \right)^2 \square} \Big]\ , \nonumber\\
&&C_3 = \int_0^1 du \Big\{ \frac{4 m_c^2 f_V m_V^3}{\triangle^3} I_2[\psi_4^\parallel - 3 \phi_3^\perp + \phi_2^\parallel](u) - 4 f_V^\perp m_c \Big[ \frac{m_V^2}{4 \triangle^2} \left( I_1[\psi_4^\perp - \phi_2^\perp](u) + \tilde{\psi}_3^\parallel(u) \right) \non
&& \hspace{1.2cm} + \left( 1 - \frac{m_c^2 -q^2 +u^2m_V^2}{\triangle} \right) \frac{m_V^2}{u \triangle^2} I_2[\phi_3^\parallel - \frac{\phi_2^\perp}{2} - \frac{\psi_4^\perp}{2}](u) \Big] \Big\}\ ,\nonumber\\
&&C_4 = \int_0^1 du \Big\{ \left(m_c^2 f_V m_V \right) \Big[ \frac{I_1[\phi_2^\parallel - \phi_3^\perp](u) }{\triangle^2} 
+ \frac{2 m_V^2 u}{\triangle^3} I_2[\psi_4^\parallel - 2 \phi_3^\perp + \phi_2^\parallel](u) \non
&& \hspace{1.2cm} - \frac{3m_c^2 m_V^2}{2 \triangle^4} I_1[\phi_4^\parallel - \phi_5^\perp](u) \Big] 
+ \left(f_V^\perp m_c \right) \Big[ \frac{ \phi_2^\perp(u)}{2 \triangle} \non 
&& \hspace{1.2cm} - \frac{m_V^2 u}{2 \triangle^2} I_1[\psi_4^\perp - \phi_2^\perp](u) 
- \frac{m_V^2 u}{2 \triangle^2} \tilde{\psi}_3^\parallel(u) - \frac{m_V^2}{8} \left( \frac{1}{\triangle^2} + \frac{2m_c^2}{\triangle^3} \right) \phi_4^\perp(u) \non
&& \hspace{1.2cm} - \left( \frac{3 m_V^2}{\triangle^2} + \frac{2 m_V^2 \left( m_c^2 -q^2 + u^2 m_V^2 \right)}{\triangle^3} \right) 
I_2[\phi_3^\parallel - \frac{\phi_2^\perp}{2} - \frac{\psi_4^\perp}{2} ](u) \Big] \Big\} \non
&& \hspace{0.6cm} + f_V m_V^3 m_c^2  \int {\cal D}\alpha_i 
\Big[ \frac{4 \left( \Phi({\bar \alpha}_1, \alpha_3) - \tilde{\Phi}({\bar \alpha}_1, \alpha_3) \right)}{\square^3} \non
&& \hspace{1.2cm} + \frac{6}{\left( \alpha_1 + v \alpha_3 \right)} \left( 1 + \frac{m_c^2 -q^2 + \left( \alpha_1 + v \alpha_3 \right)^2 m_V^2}{\square} \right) 
\frac{\Psi({\bar{\bar \alpha}}_1, \alpha_3) - \tilde{\Psi}({\bar{\bar \alpha}}_1, \alpha_3) }{\square^3} \non
&& \hspace{1.2cm} - \frac{4v}{m_V m_c \left(\alpha_1 + v \alpha_3 \right) } 
\left( 1 + \frac{m_c^2 -q^2 + \left( \alpha_1 + v \alpha_3 \right)^2 m_V^2}{\square} \right) \frac{{\cal T}({\bar \alpha}_1, \alpha_3)}{\square^2} \Big] \label{eq:CF-Ci}\ ,\eeq
with the integration measure, involved in three-particle contribution, being
\begin{align}
    \int {\cal D}\alpha_i \equiv \int_0^1 du \int_0^u d \alpha_1 \int_0^{\bar u} \frac{d \alpha_2}{1-\alpha_1 - \alpha_2}\ ,
\end{align}
In Eq.~\eqref{eq:CF-Ci}, the information of large virtuality is encoded in the notations $\triangle \equiv -\left(q+ u p \right)^2 + m_c^2$ and $\square \equiv \left[q+ \left( \alpha_1 + v \alpha_3 \right) p \right]^2 - m_c^2$ with the coordinate fraction $v = (u - \alpha_1)/(1-\alpha_1 - \alpha_2)$. For clarity, we collect all the involved LCDAs of different twists, together with the corresponding Dirac structures, in Table~\ref{tab:LCDAs-twists.2p} and Table~\ref{tab:LCDAs-twists.3p} for two- and three-particle configutations, repectively. 
We have also introduced the following auxiliary distribution functions:
\beq &&I_1[\phi](u) = \int_0^u du' \phi(u')\ , \qquad I_2[\phi](u)=\int_0^u du'\int_0^{u'} du'' \phi(u'')\ ,\non
&&\Phi({\bar \alpha}_1, \alpha_3) = \int_0^{\alpha_1} d \alpha'_1 \Phi(\alpha'_1,\alpha_3)\ , \qquad
\Psi({\bar{\bar \alpha}}_1, \alpha_3) = \int_0^{\alpha_1} d \alpha'_1 \int_0^{\alpha'_1} d \alpha^{''}_1 \Phi(\alpha''_1,\alpha_3)\ . \eeq

%---------------------------------------------------------
\begin{table}[b] 
\caption{Two-particle LCDAs for vector meson. The twists and relevant Dirac structures are also specified. Explicit expressions of the LCDAs are indicated in the last row.} \vspace{-4mm}
\begin{center} 
\setlength{\tabcolsep}{1mm}{
\begin{tabular}{c | c c | c c c c | c c c c | c c} \hline
${\rm LCDAs}$ & $\phi_2^\parallel$ & $\phi_2^\perp$ & $\phi_3^\parallel$ & $\psi_3^\parallel$ & $\phi_3^\perp$ & $\psi_3^\perp$ & $\phi_4^\parallel$ & $\psi_4^\parallel$ & $\phi_4^\perp$& $\psi_4^\perp$ & $\phi_5^\perp$ & $\psi_5^\perp$  \nonumber\\ \hline
${\rm Twist}$ & 2 & 2 & 3 & 3 & 3 & 3 & 4 & 4 & 4 & 4 & 5 & 5 \nonumber\\ \hline
${\rm Dirac \, structure}$ & $\gamma_\mu$ & $\sigma_{\mu\nu}$ & $\sigma_{\mu\nu}$ & ${\bf 1}$ & $\gamma_\mu$ & $\gamma_\mu \gamma_5$ & $\gamma_\mu$ & $\gamma_\mu$ & $\sigma_{\mu\nu}$ & $\sigma_{\mu\nu}$ & $\gamma_\mu $ & $\gamma_\mu \gamma_5$ \nonumber\\ \hline
${\rm Expression}$ & {\rm (\ref{eq:phi2})} & {\rm (\ref{eq:phi2})} & {\rm (\ref{eq:phi3para})} & {\rm (\ref{eq:psi3para})} & {\rm (\ref{eq:phi3perp})} & {\rm (\ref{eq:psi3perp})} & {\rm (\ref{eq:phi4})} & {\rm (\ref{eq:psi4})} & {\rm (\ref{eq:phi4})} & {\rm (\ref{eq:psi4})} & {\rm (\ref{eq:phi5})} & {\rm (\ref{eq:phi5})}  \nonumber\\ \hline
\end{tabular}} \end{center} \label{tab:LCDAs-twists.2p}
\end{table} 
%---------------------------------------------------------
\begin{table}[t]
 \caption{Three-particle LCDAs for vector meson. The twists and relevant Dirac structures are also specified. Explicit expressions of the LCDAs are indicated in the last row.} \vspace{-4mm}
\begin{center} 
\setlength{\tabcolsep}{2mm}{\begin{tabular}{c | c c c | c c c c c c }\hline
${\rm LCDAs}$ & ${\cal V}$ & ${\cal A}$ & ${\cal T}$ & ${\cal S}$ & $\tilde{\cal S}$ & $T_1^{(4)}$ & 
$T_2^{(4)}$ & $T_3^{(4)}$ & $T_4^{(4)}$ \nonumber\\ \hline
${\rm Twist}$ & 3 & 3 & 3 & 4 & 4 & 4 & 4 & 4 & 4 \nonumber\\ \hline 
${\rm Dirac \, structure}$ & $\gamma_\mu$ & $\gamma_\mu \gamma_5$ & $\sigma_{\mu\nu}$ & ${\bf 1}$ & $\gamma_5$ & $\sigma_{\mu\nu}$ & $\sigma_{\mu\nu}$ & $\sigma_{\mu\nu}$ & $\sigma_{\mu\nu}$ \nonumber\\ \hline
${\rm Expression}$ & {\rm (\ref{eq:LCDAs3pt3})} & {\rm (\ref{eq:LCDAs3pt3})} & {\rm (\ref{eq:LCDAs3pt3})} & {\rm (\ref{eq:LCDAs3pt4})} & {\rm (\ref{eq:LCDAs3pt4})} & {\rm (\ref{eq:LCDAs3pt4})} & {\rm (\ref{eq:LCDAs3pt4})} & {\rm (\ref{eq:LCDAs3pt4})} & {\rm (\ref{eq:LCDAs3pt4})}
\nonumber\\ \hline
\end{tabular}} \end{center} \label{tab:LCDAs-twists.3p}
\end{table}
%---------------------------------------------------------

We have checked that, at leading twist and two-particle twist-three level, our OPE results are consistent with the previous LCSR predictions in Refs.~\cite{Wu:2006rd,Fu:2020vqd}.
However, differences occur in the higher-twist terms. These differences can be understood as follows. 
Firstly, the calculation in Ref.~\cite{Wu:2006rd} was performed within the leading-power approximation of heavy quark effective field theory, thereby neglecting ${\cal O}(1/m_c)$ correction terms. Secondly, in Ref.~\cite{Fu:2020vqd}, the FFs were derived directly in the helicity basis, where correlation function is multiplied by the polarization vector $\eta^\ast$. The multiplication of $\eta^\ast$ eliminates certain terms that would contribute to the standard correlation function defined in Eq.~(\ref{eq:CF-Ci}). Finally, our calculation includes additional contributions from two-particle twist-five and three-particle LCDAs.

\subsection{LCSR results of form factors}
As the invariant variables $q^2$ and $(p+q)^2$ evolve from kinematic regions well below the threshold to the physical region near the threshold, say $q^2 \sim (p+q)^2 \sim {\cal O}(m_c^2)$, the currents in the correlation function start to produce $c$-flavored hadrons. In this region, the analyticity of the correlation function in the variable $(p+q)^2$ allows us to express the invariant densities $C[{\cal V}_i]$ at a fixed $q^2$ via a dispersion relation 
\beq C[{\cal V}_i](q^2, (p+q)^2) = \frac{1}{\pi} \int_{m_c^2}^\infty ds \frac{{\rm Im} \, C[{\cal V}_i](q^2,s)}{s-(p+q)^2}. 
\label{eq:DR} \eeq
For convenience, we have chosen to use the FF densities defined in the orthogonal basis~\eqref{eq:orth-momenta}, 
\beq C[\Gamma^\mu] = C[{\cal V}_1] p_1^\mu + C[{\cal V}_2] p_2^\mu + C[{\cal V}_3] p_3^\mu + C[{\cal V}_P] p_P^\mu. \label{eq:CF-OPE-orthogonal} \eeq 
The densities $C[{\cal V}_i]$ are readily obtainable from $C_i$ ($i=1,2,3,4$) in Eq.~\eqref{eq:CF-Ci} via
\beq &&C[{\cal V}_1] = C_1, \qquad 
C[{\cal V}_3] = - \frac{m_D^2 - m_V^2}{q^2} \left( C_4 + C[{\cal V}_2] \right), \non
&&C[{\cal V}_2] = C_2, \qquad 
C[{\cal V}_P] = C_3 - C_4 - C[{\cal V}_3]. \label{eq:relation-FFs-densities} \eeq 

A complete set of hadron states with $D_{(s)}$ quantum number are generated between the currents $\Gamma^\mu$ and $\Gamma_5$. Isolating the contribution from the lightest hadronic state $D_{(s)}$, imposing the semi-local quark-gluon duality and performing the Borel transformation, we ultimately obtain the LCSR results of $D \to V$ transition form factors, 
\beq F_i(q^2) = \kappa_{F_i} \frac{e^{m_{D_{(s)}}^2/M^2}}{m_{D_{(s)}}^2 f_{D_{(s)}}} {\hat B}\left[ C[{\cal V}_i] \right]\ ,\quad F_i \in \{V, A_0, A_1, A_3 \}\ . \label{eq:FFs-LCSRs} 
\eeq
with the prefactors given by 
\beq \kappa_V = - \left(m_{D_{(s)}} + m_V \right), \quad 
\kappa_{A_1} = - \left(m_{D_{(s)}} - m_V \right), \quad 
\kappa_{A_3} = \frac{q^2}{2m_V}, \quad 
\kappa_{A_0} = - \frac{q^2}{2m_V}. \eeq
The symbol $\hat{B}$ implies that the Borel transformation should be done. It can be easily noticed that $C[{\cal V}_i]$ (c.f. Eq.~\eqref{eq:CF-Ci}) are expressed in terms of two types of integrals, 
\beq &&{\cal I}_n = \int_0^1 du \frac{{\cal F}(u)}{\triangle^n} = (-1)^n \int_0^1 du \frac{{\cal F}(u)}{ (-\triangle)^n} 
=  (-1)^n\int_0^1 \frac{du}{u^n} \frac{{\cal F}(u)}{ \left[s_2(u,q^2) - (p+q)^2 \right]^n}\ , \non
&&{\cal I}'_n = \int {\cal D}\alpha_i \frac{{\cal F}'(u)}{\square^n} = (-1)^n\int {\cal D}\alpha_i \frac{{\cal F}'(u)}{(-\square)^n} 
= (-1)^n \int \frac{{\cal D}\alpha_i}{\left( \alpha_1 + v \alpha_3\right)^n} \frac{{\cal F}'(u)}{ \left[s_3(\alpha_i,q^2) - (p+q)^2\right]^n}\ .
\eeq
The off-shell functions involved in two-particle and three-particle contributions are 
\beq && s_2(u,q^2) = \frac{m_c^2 - {\bar u}q^2 + u {\bar u}k^2}{u}\ , \non 
&& s_3(\alpha_i,q^2) = \frac{m_c^2 - (1- \alpha_1 - v \alpha_3)q^2 + (\alpha_1 + v \alpha_3)(1- \alpha_1 - v \alpha_3)k^2}{\alpha_1 + v \alpha_3}\ . \eeq
It should be noted that, in Eq.~\eqref{eq:FFs-LCSRs}, the subtracted Borel transformation ${\hat B}$ is regarded as a two-step operation: the Borel transformation followed by the so-called continuum subtraction. The Borel transformation of ${\cal I}_1$ reads
\beq &&{\hat B}[{\cal I}_1] = {\hat B}\left[ (-1) \int_0^1 \frac{du }{u} \frac{{\cal F}(u)}{ \left[s_2(u,q^2) - (p+q)^2 \right]} \right] 
= (-1) \int_{u_0}^1 \frac{du}{u} {\cal F}(u) e^{-\frac{s_2(u,q^2)}{M^2}}\ , \eeq 
with the Borel mass $M^2$ and the subtraction threshold $s_0$. The threshold momentum fraction $u_0$ is defined as the solution to the equation $s_2(u,q^2) = s_0$. 
Transformation for ${\cal I}_{n \geqslant 1}$ can be recursively obtained via
\beq {\hat B}[{\cal I}_{n}] = \frac{1}{\Gamma[n]} \left( - \frac{d}{dm_c^2} \right)^{(n-1)} {\hat B}[{\cal I}_1]. \eeq

\section{Numerical results and discussions}\label{ffs-numerics}

\subsection{Parameter setup}

%---------------------------------------------------------
\begin{table*}[b]
\caption{Nonperturbative parameters in two- and three-particle LCDAs for vector mesons.}
\begin{center} \vspace{-4mm}
\setlength{\tabcolsep}{1.5mm}{
\begin{tabular}{c | c c c c c c c c } \hline
${\bf Parameters}$ & $a_1^\parallel$ & $a_1^\perp$ & $a_2^\parallel$ & $a_2^\perp$ & $\zeta_3^\parallel$ & $\omega_3^\parallel$ & $\omega_3^\perp$ & $\tilde{\omega}_3^\parallel$ \nonumber\\ \hline
 ${\rho}$ & - & - & $0.17(7) $ & $0.14(4)$ & $0.03(1)$ & $0.15(5)$ & $0.55(25)$ & $-0.09(3)$ \nonumber\\ %\hline
${K^\ast}$ & $0.06(4)$ & $0.04(3)$ & $0.16(9) $ & $0.10(8)$ & $0.02(1)$ & $0.10(4)$ & $0.3(1)$ & $-0.07(3)$ \nonumber\\ \hline %\end{tabular} } \\ \vspace{4mm}
${\bf Parameters}$ & $\zeta_3$ & $\omega_3^V$  & $\omega_3^A$ & $\omega_3^T$ & $\zeta_4$ & $\zeta_4^T$ & $\tilde{\zeta}_4^T$  \nonumber\\ \hline
${\rho}$ & $0.03(1)$ & $3.3(1.1)$ & $-6.0(2.0) $ & $12(5.5)$ & $0.15(5)$ & $0.10(5)$ & $-0.10(5)$  \nonumber\\ %\hline 
${K^\ast}$ & $0.02(1)$ & $2.9(1.2)$ & $-6.0(2.1) $ & $8.7(2.9)$ & $0.15(5)$ & $0.10(5)$ & $-0.10(5)$  \nonumber\\ \hline \end{tabular} } \\ 
\end{center} \label{tab:LCDAs-paras}
\end{table*} 
%---------------------------------------------------------

In our numerical computation, the masses of the $\rho$ and $K^\ast$ mesons are set to be $m_\rho=775$ MeV and $m_{K^\ast}=892$ MeV~\cite{ParticleDataGroup:2024cfk}, respectively. The values of charmed meson masses are: $m_D =1.86$ GeV, $m_{D_s} =1.97$ GeV~\cite{ParticleDataGroup:2024cfk}. 
In addition, the decay constants of charmed mesons are chosen as $f_D = 0.208$ GeV~\cite{Kuberski:2024pms} and $f_{D_s} = 0.251$ GeV~\cite{BESIII:2021bdp}.  Following Ref. \cite{Bharucha:2015bzk}, the longitudinal decay constants are determined, $f_\rho^\parallel = 0.22(1)$~GeV and $f_{K^\ast}^\parallel =0.20(1)$~GeV, by using experimental measurements of the charged and neutral decay modes of $V^0 \to e^+e^-$ and $\tau^+ \to V^+ \nu$. For the scale-dependent transverse decay constants $f^\perp_V(\mu)$, we extract them from lattice simulation results of the ratios $r_V(\mu) = f_V^\perp(\mu)/f_V^\parallel$~\cite{RBC-UKQCD:2008mhs}. Setting the renormalization scale at $\mu=1$~GeV, we get $f_\rho^\perp = f_{K^\ast}^\perp = 0.16(1)$ GeV. There are a multitude of nonpertubative parameters appearing in the LCDAs of the vector mesons, whose values are collected in table~\ref{tab:LCDAs-paras}. The Gegenbauer coefficients $a_i^{\parallel (\perp)}$ (see e.g. Eq.~\eqref{eq:phi2}) are pinned down by a combined analysis of the lattice simulation and QCD sum rule calculations~\cite{Bharucha:2015bzk,Arthur:2010xf}. 
We take the twist-three parameters from Ref.~\cite{Ball:2007rt}, where the relations between parameters entering the two-particle and three-particle LCDAs, say $\omega_3^\parallel = \frac{3}{2} \zeta_3^\parallel \omega_3^V$, $\tilde{\omega}_3^\parallel = \frac{1}{2} \zeta_3^\parallel \omega_3^A$ and $\omega_3^\perp = \frac{3}{2} \zeta_3^\parallel \omega_3^T$, are considered.
And, for the twist-four parameters $\zeta_4^T, \tilde{\zeta}_4^T, \zeta_4$, we follow the determinations in Ref.~\cite{Wu:2006rd}. Note that, for the scale-dependent parameters, the default value of the renormalization scale is taken again at $1$ GeV. The equations of scale evolution and anomalous dimensions are listed in Appendix~\ref{app:LCDAs}. For charmed meson decays, the characteristic energy scale is usually chosen as $\mu = \sqrt{m_{D_{(s)}}^2 - m_c^2} \simeq 1.3 \, \text{GeV}.$ 

In a LCSR calculation, the Borel transformation and quark-hadron duality introduce two extra parameters, the Borel mass squared $M^2$ and continuum threshold $s_0$, to isolate the ground-state contribution. The two parameters are process-dependent, and the varying of their values accounts for the systematical errors of the LCSR approach, to be discussed in the following. 
The Borel mass squared characterizes the virtuality of the internal quark propagator $M^2 \sim \mathcal{O}(u m_{D}^2 + \bar{u} q^2 - u\bar{u} m_V^2)$. As a first estimate, the $M^2$ parameter should smaller than the continuum threshold, and, meanwhile, slightly larger than the factorization scale, ie., $\mu^2 \leqslant M^2 \le s_0$. In practice, the optimal Borel mass is determined through a physically motivated compromise in the sum rule analysis. This balance ensures two key requirements. On the one hand, in the hadronic representation, the ground-state contribution remains overwhelmingly dominant, which favors a smaller value of $M^2$. On the other hand, the OPE maintains good convergence at the quark-gluon level, which prefers a larger one. Mathematically, this dual constraint leads to an extremization procedure within a conventional parameter range, say $\frac{d}{d(1/M^2)} \ln F_{i}(q^2) = 0$.
The threshold parameter $s_0$ is typically chosen to be a value around the mass of first excited state of the interpolating hadron, and justified by the stability of physical quantities as the Borel mass squared varies. According to the above-mentioned requirements, an empirical selection criterion is to ensure that the contribution of excited states and the continuum in the hadronic representation does not exceed \( 30\% \). Besides, the high-twist contributions in the OPE at the quark-gluon level remain smaller than the leading-twist one. Eventually, we set Borel mass squared $M^2$ at $4.5 \pm 1$ GeV$^2$ and $4.0 \pm 1.0$ GeV$^2$ for the $D \to \rho$ and $D_{(s)} \to K^\ast$ transitions, respectively. The continuum threshold is set to be $s_0 = 7 \pm 0.5$ GeV$^2$. 

\subsection{Form factors in the LCSR accessible region}

\begin{figure}[t]\centering \vspace{-4mm}
\centering 
\includegraphics[width=0.465\linewidth]{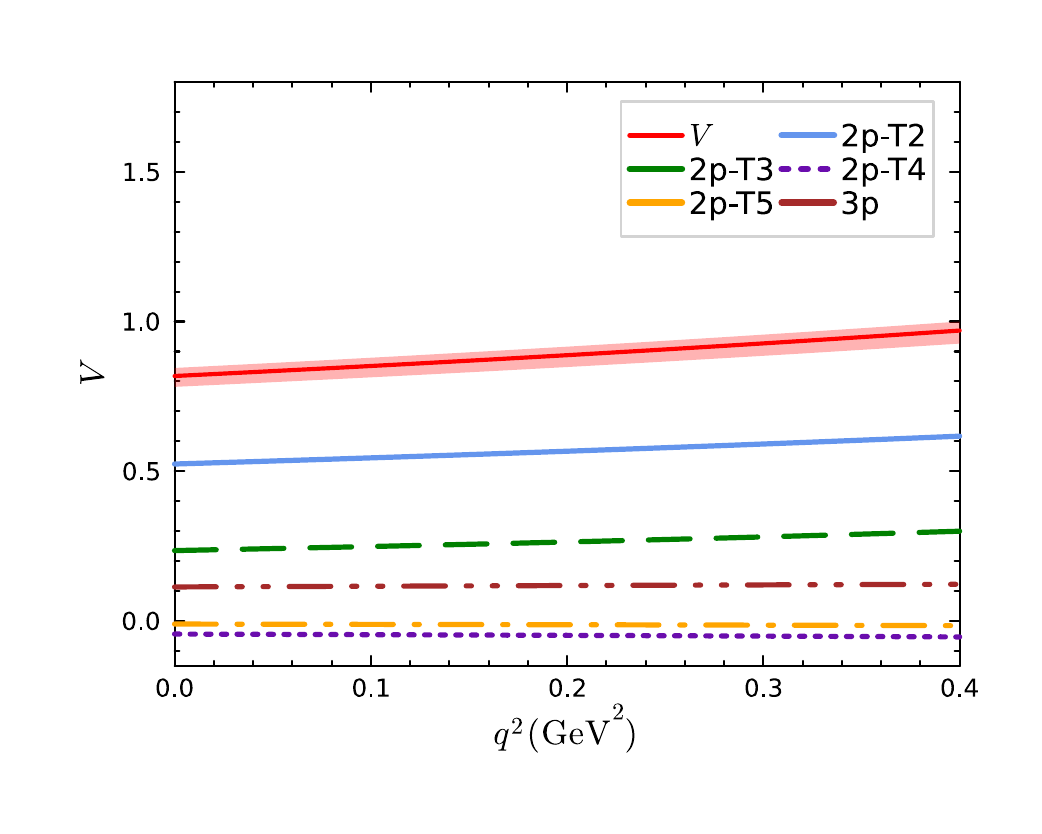}
\includegraphics[width=0.465\linewidth]{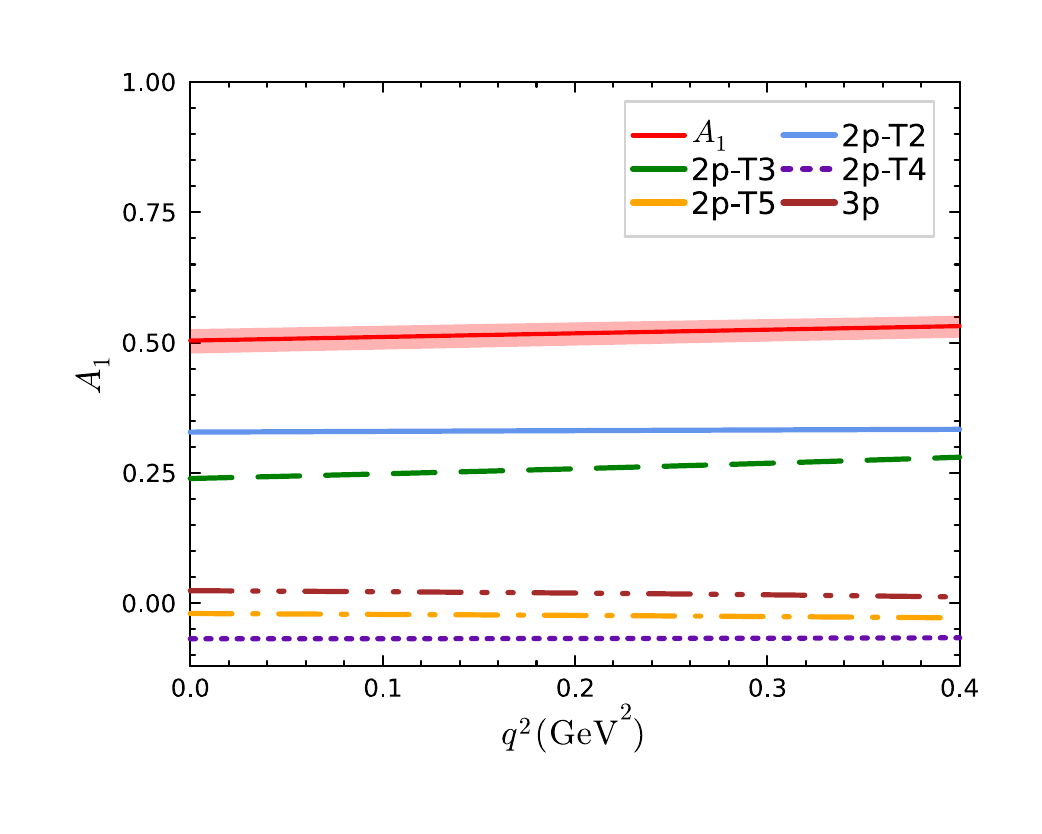} \non \vspace{-4mm}
\includegraphics[width=0.465\linewidth]{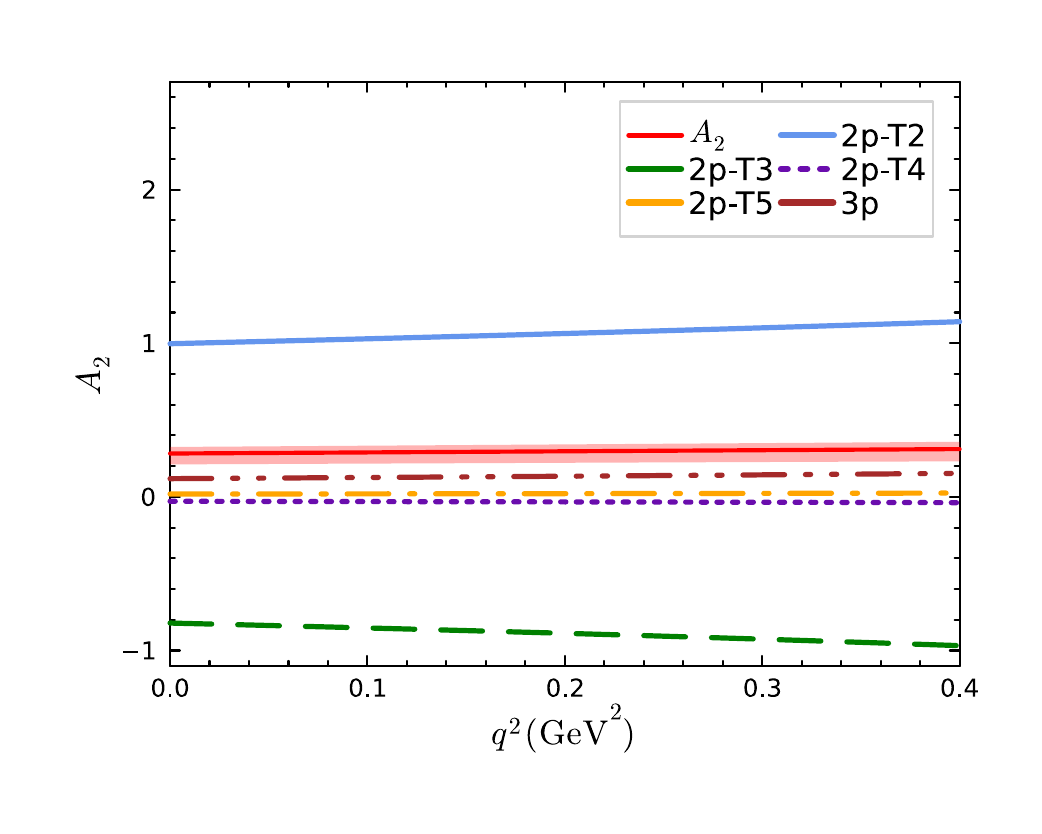}
\includegraphics[width=0.465\linewidth]{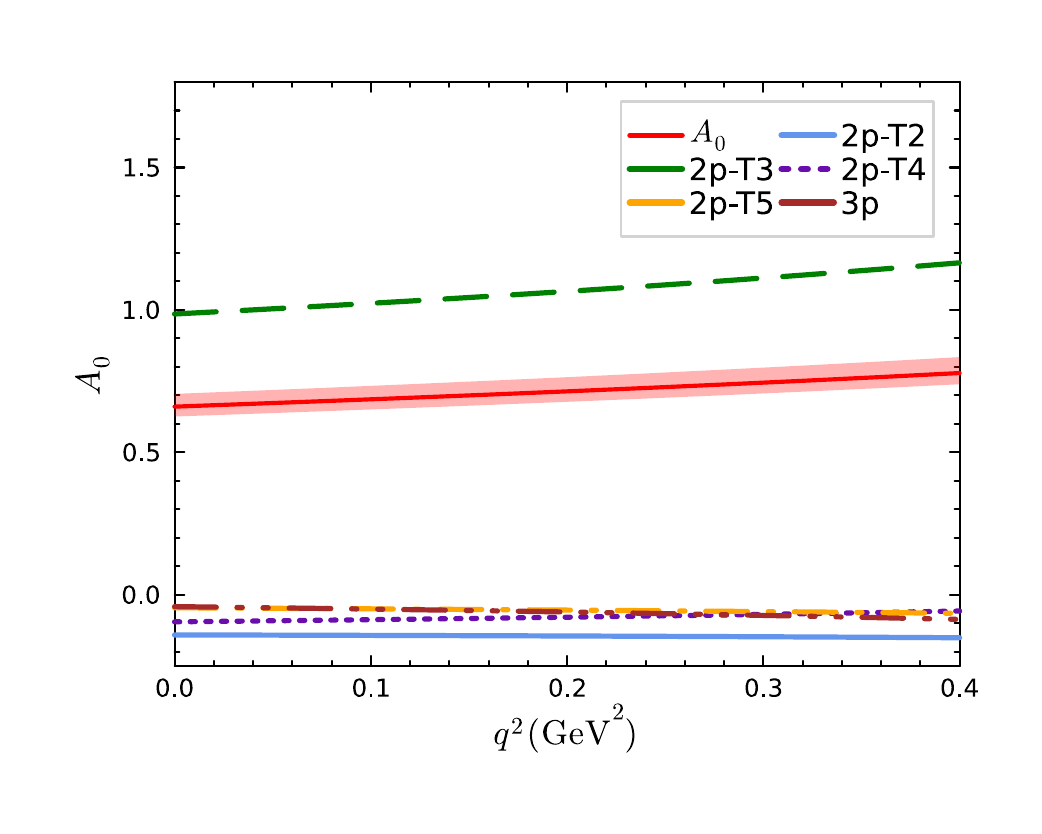} \vspace{-6mm}
\caption{LCSR predictions of the $D^0 \to \rho^-$ FFs in the region of $q^2\in [0,0.4]$~GeV$^2$. 
The FFs are anatomized into various contributions from different twist LCDAs.} 
\label{fig:D+2rho0}
\end{figure}

\begin{figure}[t]\vspace{-4mm}
\centering
\includegraphics[width=0.465\linewidth]{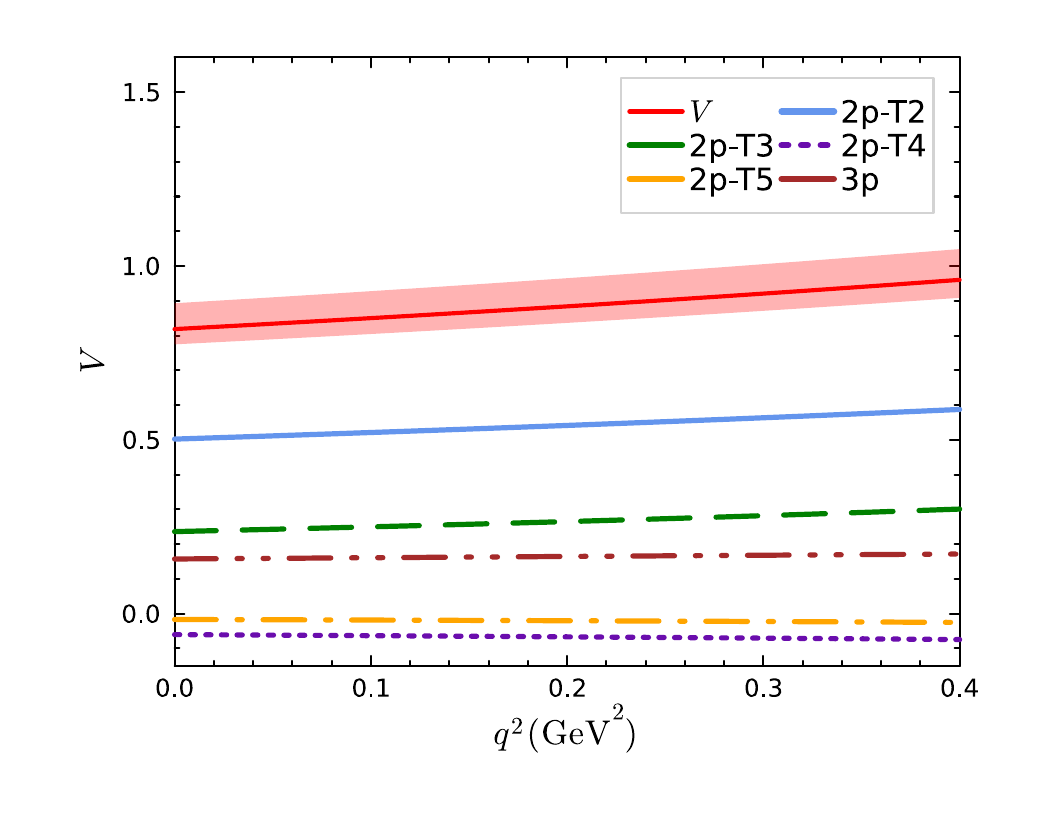}
\includegraphics[width=0.465\linewidth]{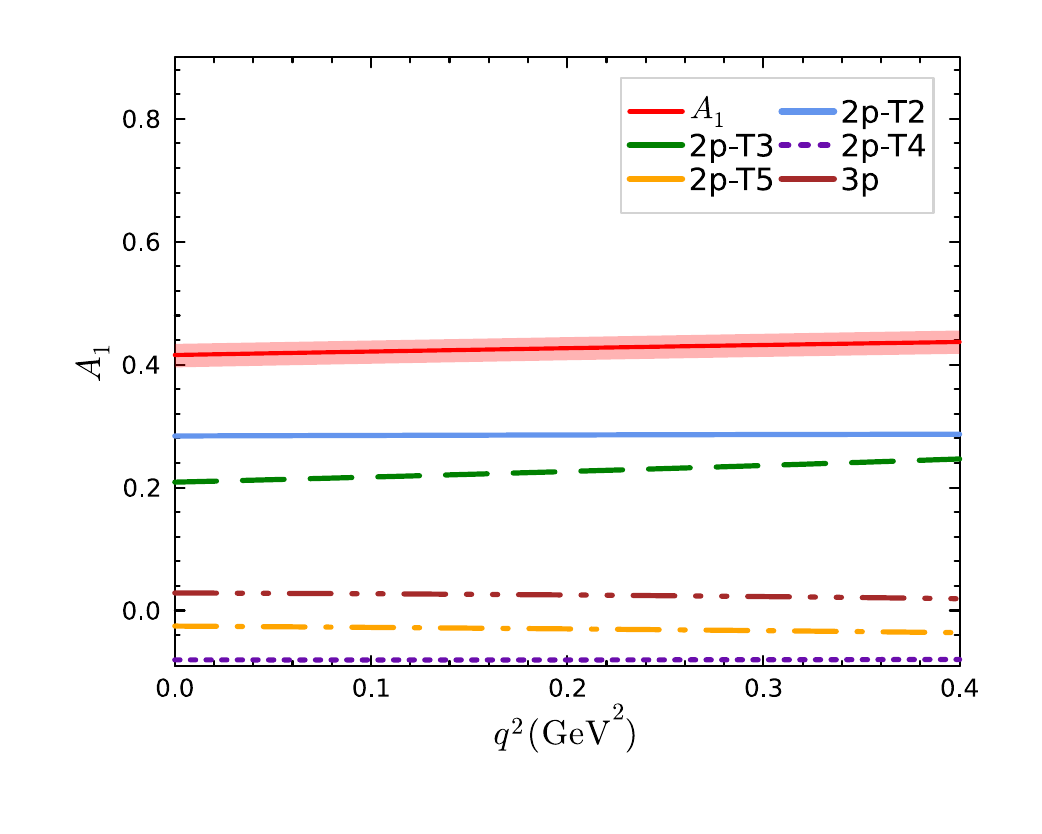} \non \vspace{-4mm}
\includegraphics[width=0.465\linewidth]{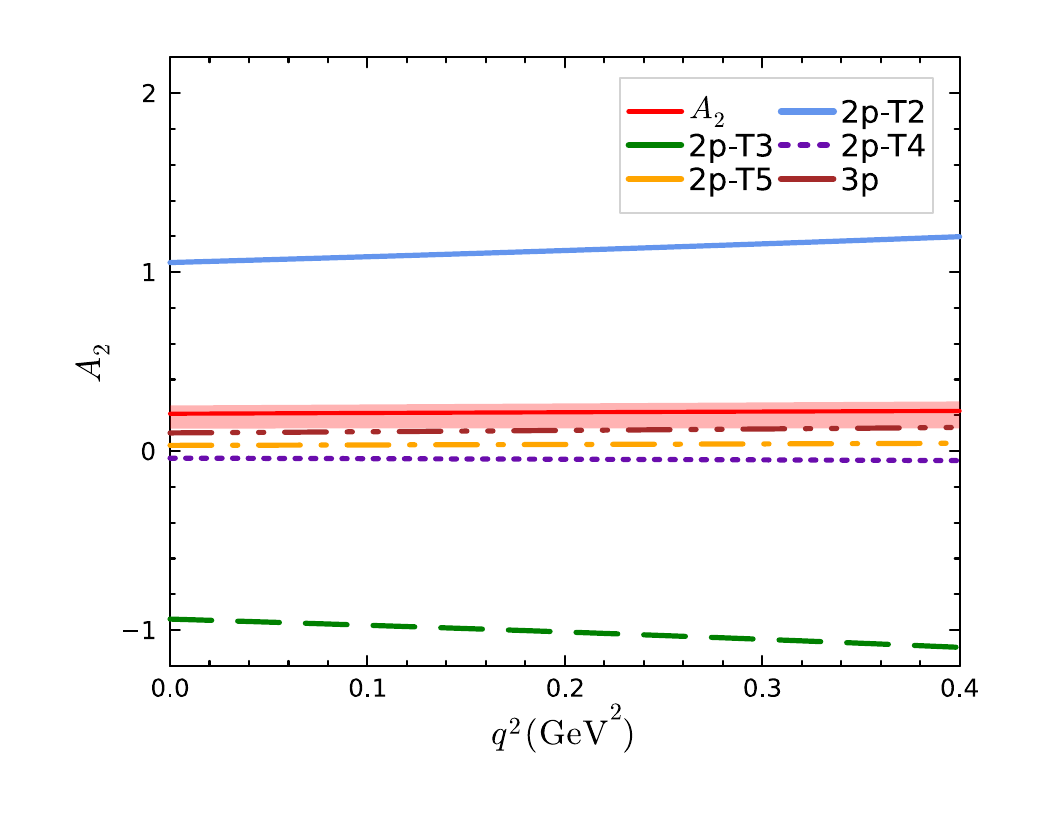}
\includegraphics[width=0.465\linewidth]{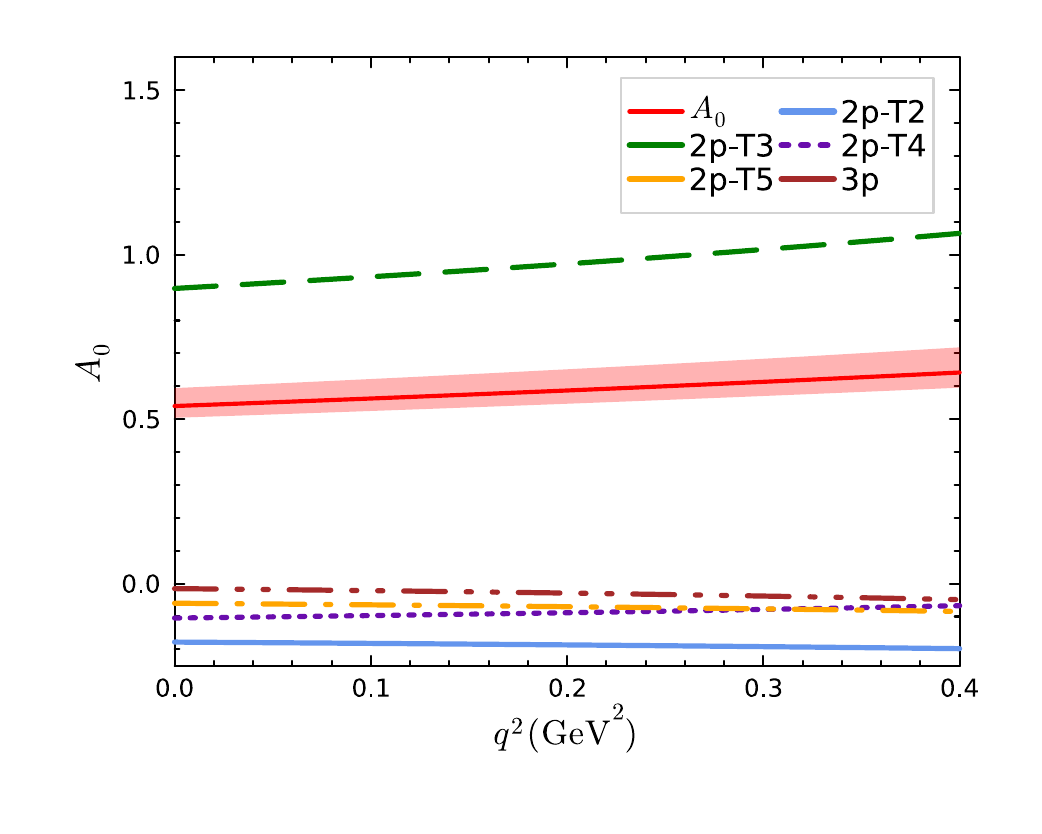} \vspace{-6mm}
\caption{LCSR predictions of $D_s^+ \to K^{\ast 0}(892)$ FFs in the region of $q^2\in [0,0.4]$~GeV$^2$. }
\label{fig:Ds+2Ks0} 
\end{figure}

\begin{figure}[t] \vspace{-4mm}
\centering \includegraphics[width=0.465\linewidth]{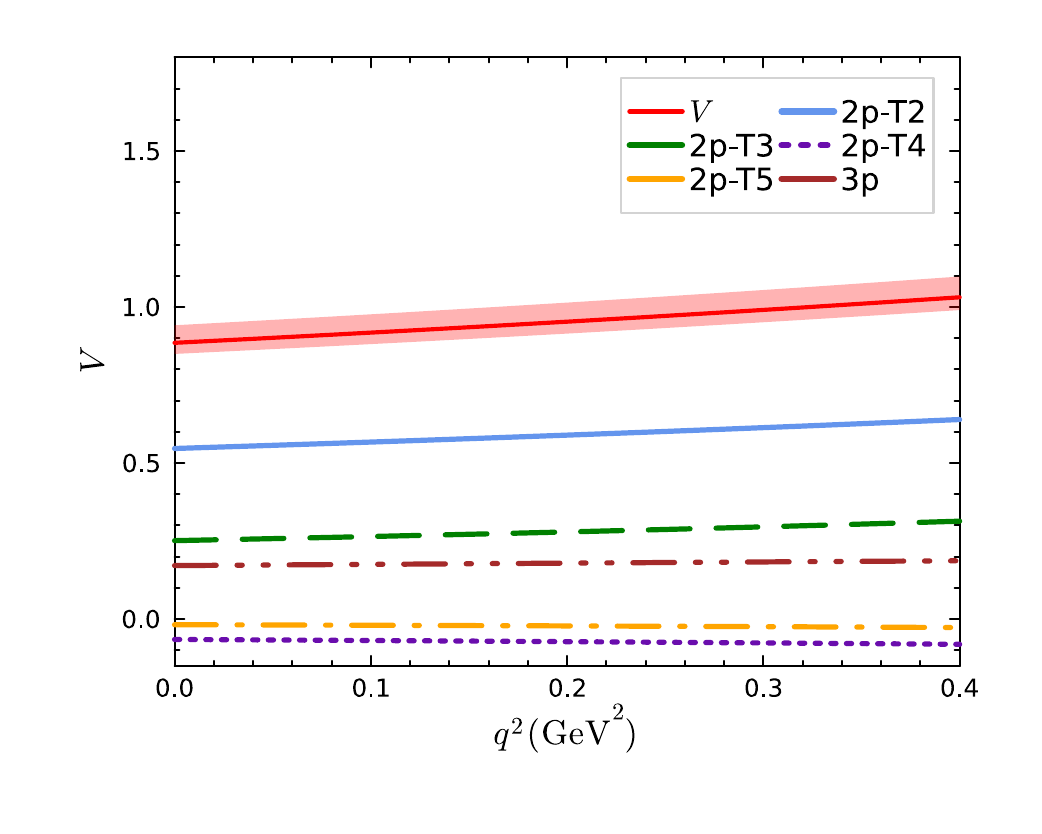}
\includegraphics[width=0.465\linewidth]{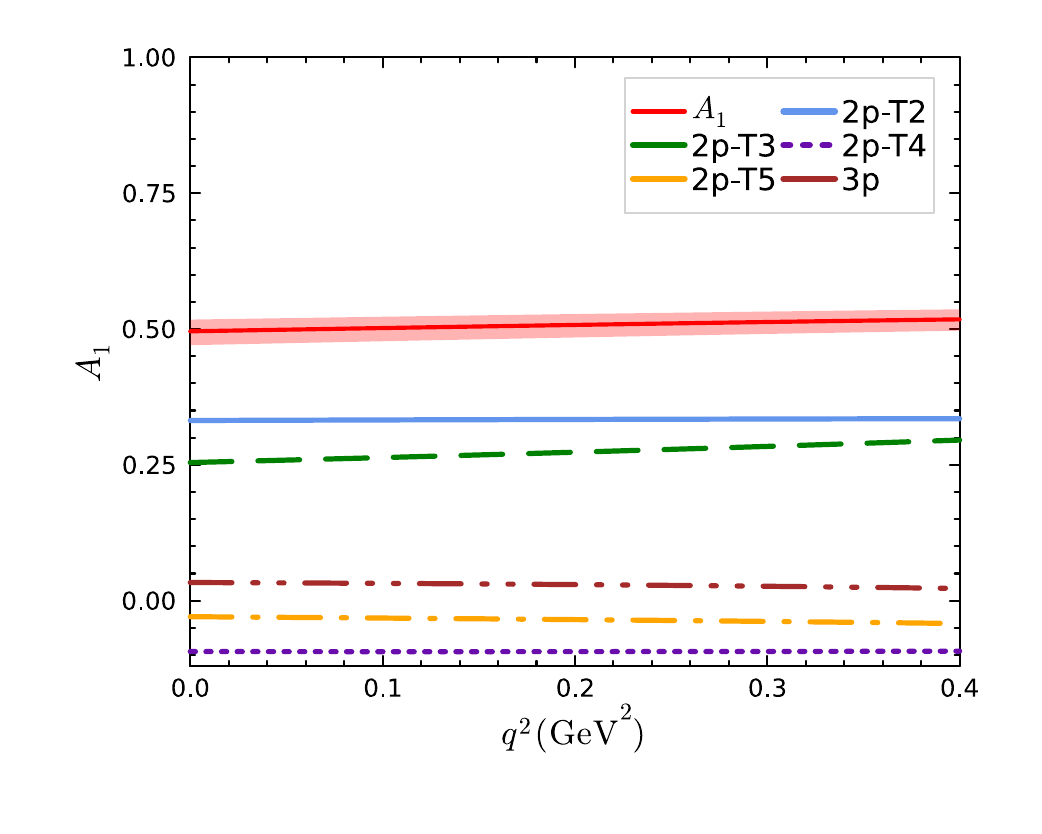}\non \vspace{-4mm}
\includegraphics[width=0.465\linewidth]{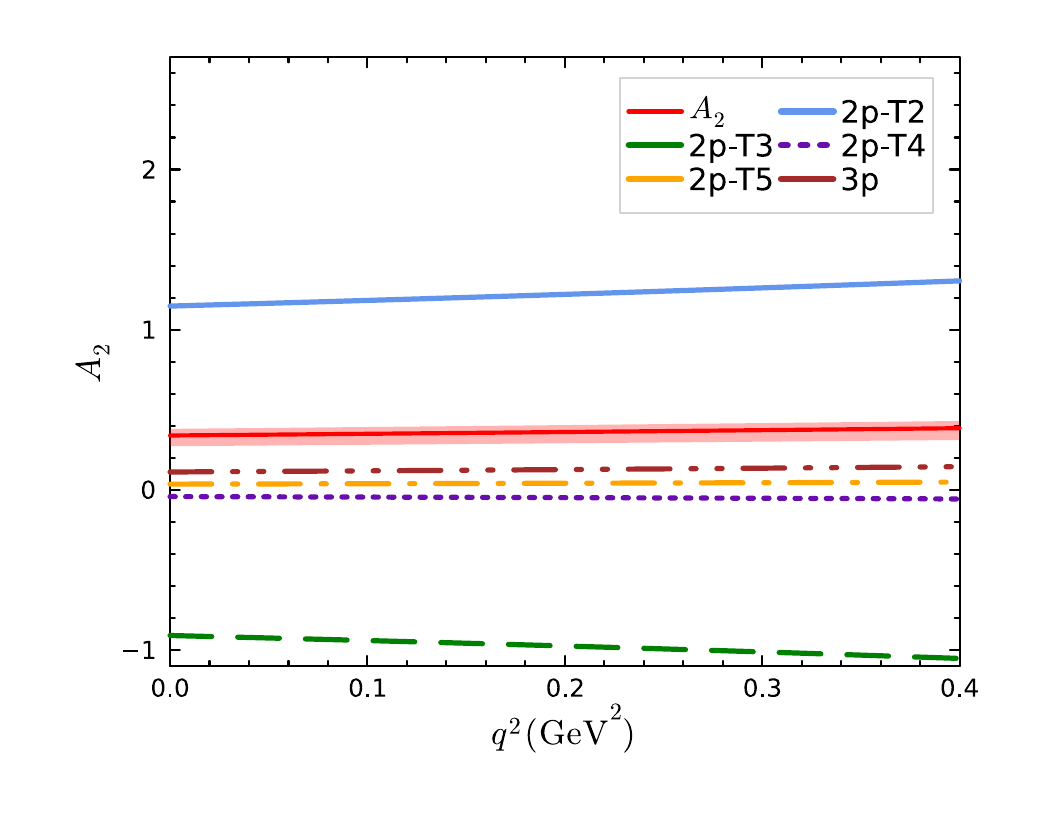}
\includegraphics[width=0.465\linewidth]{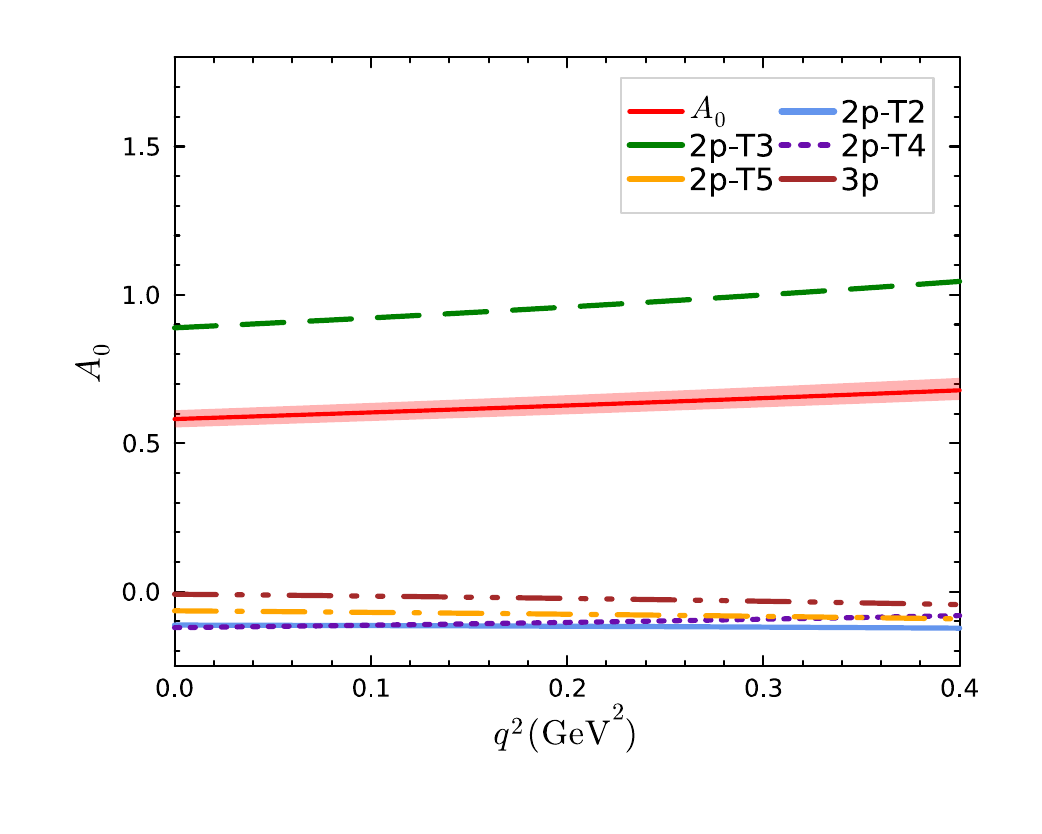} \vspace{-6mm}
\caption{LCSR predictions of $D^+ \to K^{\ast 0}(892)$ FFs in the region of $q^2\in [0,0.4]$~GeV$^2$. }
\label{fig:D+2Ks0}
\end{figure}

With the parameters discussed in the preceding subsection, we are now in the position to make predictions for the FFs. As is well known, the OPE framework is only applicable in the low $q^2$ region with large recoils. More specially, it works for momentum transfers in the range $0 \leqslant q^2 \leqslant m_c^2 - 2 m_c \chi$. This implies that the LCSRs predictions of $D_{(s)} \to V$ FFs are reliable up to $0.4$ GeV$^2$. In the region of  $q^2\in[0,0.4]$~GeV$^2$, our LCSR results of the FFs for the $D^0 \to \rho^-$, $D^+ \to K^{\ast 0}$ and $D_s^+ \to K^{\ast 0}$ transitions are plotted in Figs.~\ref{fig:D+2rho0}, \ref{fig:Ds+2Ks0} and \ref{fig:D+2Ks0}, respectively. The red solid lines with bands stand for the net LCSR results, where two-particle twist-two to -five and three-particle twist-four to -five contributions are summed. To assess the convergence property of the OPE, contributions from two-particle LCDAs of different twists are shown separately. In addition, the three-particle contributions are represented by the brown dot-dot-dashed lines. 

%---------------------------------------------------------
%---------------------------------------------------------
%---------------------------------------------------------
%---------------------------------------------------------
%---------------------------------------------------------

%---------------------------------------------------------
\begin{table}[t]\vspace{-4mm}
\caption{Form factor ratios of $D \to \rho$, $D_s \to K^\ast$ and $D \to K^\ast$ transitions.} \label{tab:ff-ratios}
\centering
\renewcommand{\arraystretch}{1.2} 
\setlength{\tabcolsep}{0.0mm}{
\begin{tabular}{c|cc cc cc} \toprule \toprule 
{\bf transitions} \quad &  \multicolumn{2}{c}{$D \to \rho $}  &  \multicolumn{2}{c}{$D_s \to K^\ast $}  & 
 \multicolumn{2}{c}{$D \to K^\ast $}  \\
{\bf ratios} \quad & \quad $r_V$ \quad & \quad $r_2$ \quad & \quad 
$r_V$ \quad & \quad $r_2$ \quad & \quad 
$r_V$ \quad & \quad $r_2$ \quad \\ \toprule
This work \quad & \quad $1.64^{+0.11}_{-0.17}$ \quad & \quad $0.56^{+0.11}_{-0.16}$ \quad & \quad $1.95^{+0.28}_{-0.16}$ \quad & \quad $0.50^{+0.13}_{-0.20}$ \quad & \quad $1.78\pm0.13$ \quad & \quad $0.68^{+0.11}_{-0.09}$ \quad \\ 
LCSR2006 \cite{Wu:2006rd} \quad & \quad $1.34^{+0.16}_{-0.13}$ \quad & \quad $0.62\pm0.08$ \quad & \quad $1.31^{+0.19}_{-0.16}$ \quad & \quad $0.53^{+0.09}_{-0.16}$ \quad & \quad $1.39^{+0.09}_{-0.10}$ \quad & \quad $0.60^{+0.09}_{-0.08}$ \quad \\ 
QM \cite{Melikhov:2000yu} \quad & \quad $1.53$ \quad & \quad $0.83$ \quad & \quad $1.82$ \quad & \quad $0.74$ \quad & \quad $1.56$ \quad & \quad $0.74$ \quad \\ 
LFQM \cite{Verma:2011yw} \quad & \quad $1.47$ \quad & \quad $0.78$ \quad & \quad $1.55$ \quad & \quad $0.82$ \quad & \quad $1.36$ \quad & \quad $0.82$ \quad \\ 
LEChQM \cite{Palmer:2013yia} \quad & \quad $1.47$ \quad & \quad $0.84$ \quad & \quad $1.53$ \quad & \quad $0.91$ \quad & \quad $1.56$ \quad & \quad $0.89$ \quad \\ 
HChPT \cite{Fajfer:2005ug} \quad & \quad $1.72$ \quad & \quad $0.51$ \quad & \quad $1.93$ \quad & \quad $0.55$ \quad & \quad $1.60$ \quad & \quad $0.50$ \quad \\ 
LQCD \cite{Abada:2002ie} \quad & \quad $1.69$ \quad & \quad $0.85$ \quad & 
\quad --- \quad & \quad --- \quad & \quad $1.47^{+0.14}_{-0.13}$ \quad & \quad $0.60\pm0.07$ \quad \\ 
CLEO \cite{CLEO:2011ab} \quad & \quad $1.48\pm0.15$ \quad & \quad $0.83\pm0.11$ \quad & \quad --- \quad & \quad --- \quad & 
\quad --- \quad & \quad --- \quad \\ 
BESIII \cite{BESIII:2024lxg,BESIII:2018xre,BESIII:2024xjf} \quad & \quad $1.55 \pm 0.09$ \quad & \quad $0.82 \pm 0.06$ \quad & \quad $1.67 \pm 0.38$ \quad & \quad $0.77 \pm 0.29$ \quad & \quad $1.48 \pm 0.06$ \quad & \quad $0.70 \pm 0.05$ \quad \\ 
\bottomrule \bottomrule 
\end{tabular}}
\end{table}
%---------------------------------------------------------

It can be found that the vector ($V$) and axial vector ($A_1$) FFs are dominated by leading-twist LCDAs, while two-particle twist-3 LCDAs also contribute significantly. Notably, the $A_1$ form factor receives additional sizable contributions from three-particle LCDAs. In contrary, higher-twist LCDAs have only minor influences on the FFs. Consequently, the collective contributions from two-particle twist-two and twist-three LCDAs produce a net positive enhancement in the final result. For the FF $A_2$, the leading-twist and two-particle twist-three LCDAs contributions are comparable in magnitude but opposite in sign. The cancellation between the two contributions leads to a significantly suppressed $A_2$ value compared to other FFs, thereby amplifying the relative importance of three-particle LCDAs contributions. 
The FF $A_0$ exhibits a markedly different behavior. Namely, the two-particle twist-three effect dominates, while the other contributions, including leading-twist, turn out to be negligible. 

The dominance of the twist-three LCDAs can be clearly explained in the framework of heavy quark effective field theory (HQEFT). In HQEFT, the matrix elements can be expanded in the power of $1/m_Q$ via 
\beq
&&\langle V(p,\eta') \vert {\bar q} \Gamma c_\nu^+ \vert {\cal D}_\nu \rangle = -i {\rm Tr}\left[{\cal F}(v,p) \Gamma {\cal D}_\nu\right], \non
&&{\cal F}(v,p) = L_1(v \cdot p) \etasl^\ast + 
L_2(v \cdot p) \left( v \cdot \eta^\ast \right) +
\left[ L_3(v \cdot p) \etasl^\ast + 
L_4(v \cdot p) \left( v \cdot \eta^\ast \right) \right] \frac{\psl}{v \cdot p}. 
\eeq
In the above expression, ${\cal D}_\nu$ is the heavy pseudoscalar spin wave function in HQEFT, $v$ is the four-velocity of heavy meson satisfying $v^2 = 1$. The FFs defined in Eq. (\ref{eq:FFs-def1}) can be recast in terms of the universal wave function $L_i(v \cdot p)$ in the heavy quark limit. Specifically, the form factor $V$ is proportional to the function $L_3$ at the leading power of heavy quark expansion, and the form factor $A_1$ related both to the functions $L_3$ and $L_1$. Interestingly, the form factors $A_2$ and $A_0$, associated with the helicity-flipped function $L_2$~\cite{Wu:2006rd}, are suppressed by powers of ${\cal O}(1/m_D)$, leading to the twist-three dominance.

At zero momentum transfer $q^2=0$, it is useful to predict the hadronic FF ratios that can be measured by experiments~\cite{CLEO:2011ab,BESIII:2018xre,BESIII:2024lxg,BESIII:2024xjf}. To that end, we define the following two ratios
\beq
r_V \equiv \frac{V(0)}{A_1(0)}, \quad r_2 \equiv \frac{A_2(0)}{A_1(0)}. 
\eeq
Our new LCSR predictions for these ratios are summarized in Table~\ref{tab:ff-ratios}. For comparison, we also show other theoretical and experimental results, including previous LCSR calculations performed in the heavy-quark limit with lower-twist LCDAs~\cite{Wu:2006rd}, quark model (QM) estimations~\cite{Melikhov:2000yu}, light front quark model (LFQM) predictions~\cite{Verma:2011yw}, large energy chiral quark model (LEChQM) result~\cite{Palmer:2013yia}, heavy meson chiral perturbation theory (HChPT) calculations~\cite{Fajfer:2005ug}, lattice QCD (LQCD) simulations~\cite{Abada:2002ie}, CLEO~\cite{CLEO:2011ab} and BESIII measurements~\cite{BESIII:2024lxg,BESIII:2018xre,BESIII:2024xjf}.  

One can see that our LCSR determinations are more or less consistent with the quoted results. We find that the ratio $r_V$ exhibits more sensitivity to ${\cal O}(1/m_c)$ corrections than $r_2$, when confronting our results with the leading-power LCSRs predictions~\cite{Wu:2006rd}. This is due to the partial cancellation of sizable power effects between the $A_2$ and $A_1$ FFs. When comparing to the recent BESIII measurements~\cite{BESIII:2024lxg,BESIII:2018xre,BESIII:2024xjf}, our calculations reveal the relatively smaller $r_2$ values in the $c \to d$ transition ($D \to \rho$ and $D_s \to K^\ast$) and significantly larger $r_V$ value in $c \to s$ transition ($D \to K^\ast$). We suggest that finite-width effect and non-resonant background in $D \to V l\nu$ decays may play an important role in reducing these observed derivations~\cite{Cheng:2025hxe}. A more detailed investigation of these effects will be presented in a followup work~\cite{D2pipi-LCSRs}. 

\subsection{Form factors in the whole kinematical region}

%---------------------------------------------------------
\begin{table*}[b] 
\caption{Masses (in unit of GeV) of the lowest excited states with quantum numbers $J^P$ and fit results of the dimensionless coefficients $a_i$ ($i=1,2,3$) in the BCL parametrization.} \label{tab:mDVR} \vspace{-4mm}
\begin{center}
\setlength{\tabcolsep}{0.1mm}{\begin{tabular}{c|c|cc|ccc|ccc|ccc} \toprule \toprule
$F_i$ \quad & $J^P$ \quad & \quad $M_{R,i}^{c\rightarrow s}$ \quad & \quad $M_{R,i}^{c\rightarrow d}$ \quad &
\makecell{~ \\ \quad $a_1$} & \quad \makecell{$D \to \rho$ \\ $a_2$} & \quad \makecell{~ \\ $a_3$} \quad &
\makecell{~ \\ \quad $a_1$} & \quad \makecell{$D_s \to K^\ast$ \\ $a_2$} & \quad \makecell{~ \\ $a_3$} \quad &
\makecell{~ \\ \quad $a_1$} & \quad \makecell{$D \to K^\ast$ \\ $a_2$} & \quad \makecell{~ \\ $a_3$} \quad \\ \midrule
$A_0$ \quad & $0^-$ & $1.97$ & $1.87$ & \quad $0.46$ & $-1.11$ & $18.4$ \quad & \quad $0.54$ & $-2.15$ & $10.3$ \quad & \quad $0.58$ & $-2.05$ & $-4.01$ \quad \\
$V$ \quad & $1^-$ & $2.12$  & $2.01$ & \quad $0.58$ & $-2.33$ & $24.2$ \quad & \quad $0.82$ & $-3.30$ & $34.0$ \quad & \quad $0.89$ & $-3.57$ & $~29.9$ \quad \\ 
$A_1$\quad & $1^+$ & $2.46$  & $2.43$ & \quad $0.36$ & $0.31$ & $-6.18$ \quad & \quad $0.42$ & $0.42$ & $-20.2$ \quad  & \quad $0.50$ & $0.61$ & $-22.1$ \quad \\
$A_2$ \quad & $1^+$ & $2.46$  & $2.43$ & \quad $0.20$ & $-0.49$ & $-7.06$ \quad & \quad $0.21$ & $-0.23$ & $-16.9$ \quad & \quad $0.34$ & $-1.45$ & $ 4.14$ \quad \\\bottomrule \bottomrule \end{tabular} }
\end{center}
\end{table*}
%---------------------------------------------------------

To extend the LCSR predictions across the entire kinematical range, $0 \leqslant q^2 \leqslant \left( m_{D_{(s)}}-m_V \right)^2$, we employ the Bourrely-Caprini-Lellouch (BCL) parametrization \cite{Bourrely:2008za}, 
\beq F_i(q^2) = P_i(q^2)\sum_k\alpha_k^i\left[z(q^2)-z(0) \right]^k. \eeq
The BCL parametrization corresponds to a combination of nearest pole dominance and a complex $z$-series expansion with 
\beq
P_i(q^2) = \frac{1}{1 - q^2/m_{R,i}^2}\ , \quad
z(q^2) = \frac{\sqrt{t_{+} - q^2} - \sqrt{t_{+} - t_{0}}}{\sqrt{t_{+} - q^2} + \sqrt{t_{+} - t_{0}}}\ ,
\eeq
where $m_{R,i}$ denotes the lowest pole in the spectrum, and the $t$-variables are defined by
\begin{align}
   t_{\pm} = (m_D \pm m_V)^2\ ,\quad t_0 = t_+(1 - \sqrt{1 - t_-/t_+}) \ .
\end{align} 
The third column of Table~\ref{tab:mDVR} shows the masses of the lowest excited states of various $J^P$ quantum numbers. In practice, the series in the BCL parametrization should be truncated at certain order. In our case, the truncation is done at the third order. The coefficients $a_i$'s are unknown constants and are fixed by fitting to the LCSR predictions. Fit results of the coefficients for $D\to \rho$, $D_s\to K^\ast$ and $D\to K^\ast$ are collected in Table~\ref{tab:mDVR}.

%---------------------------------------------------------
\begin{figure}[t]\centering \vspace{-4mm}
\centering \includegraphics[width=0.465\linewidth]{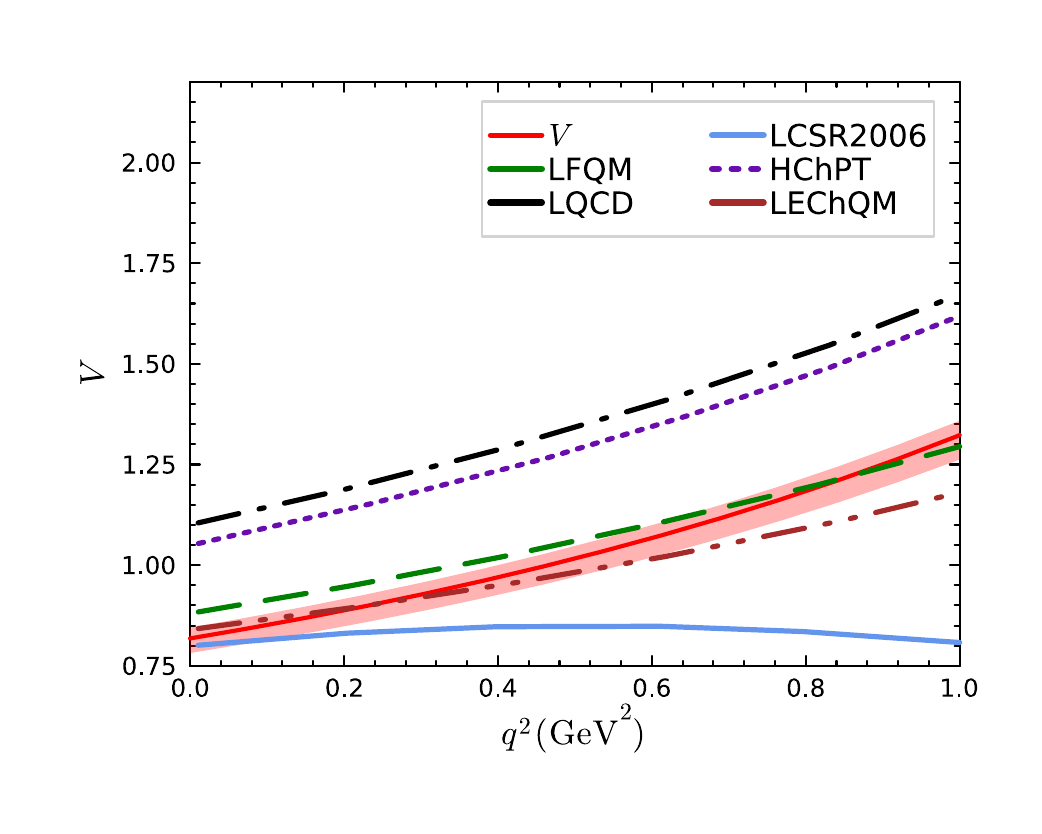}
\includegraphics[width=0.465\linewidth]{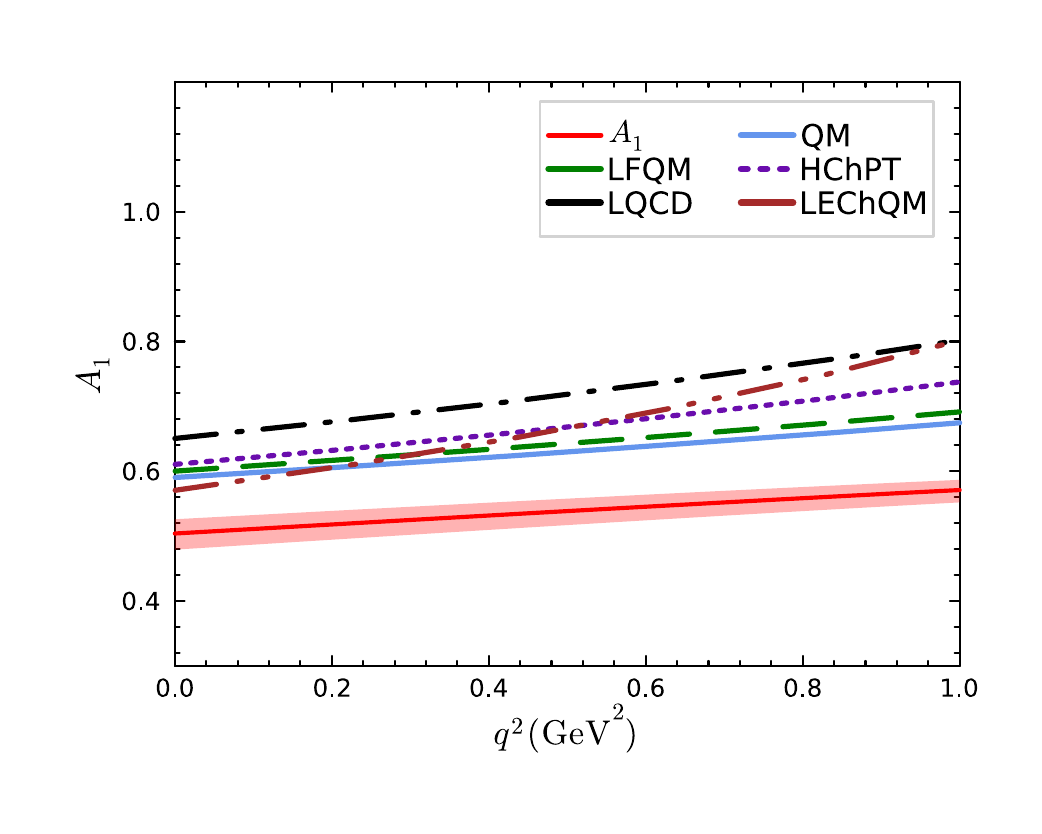}\non\vspace{-4mm}
\includegraphics[width=0.465\linewidth]{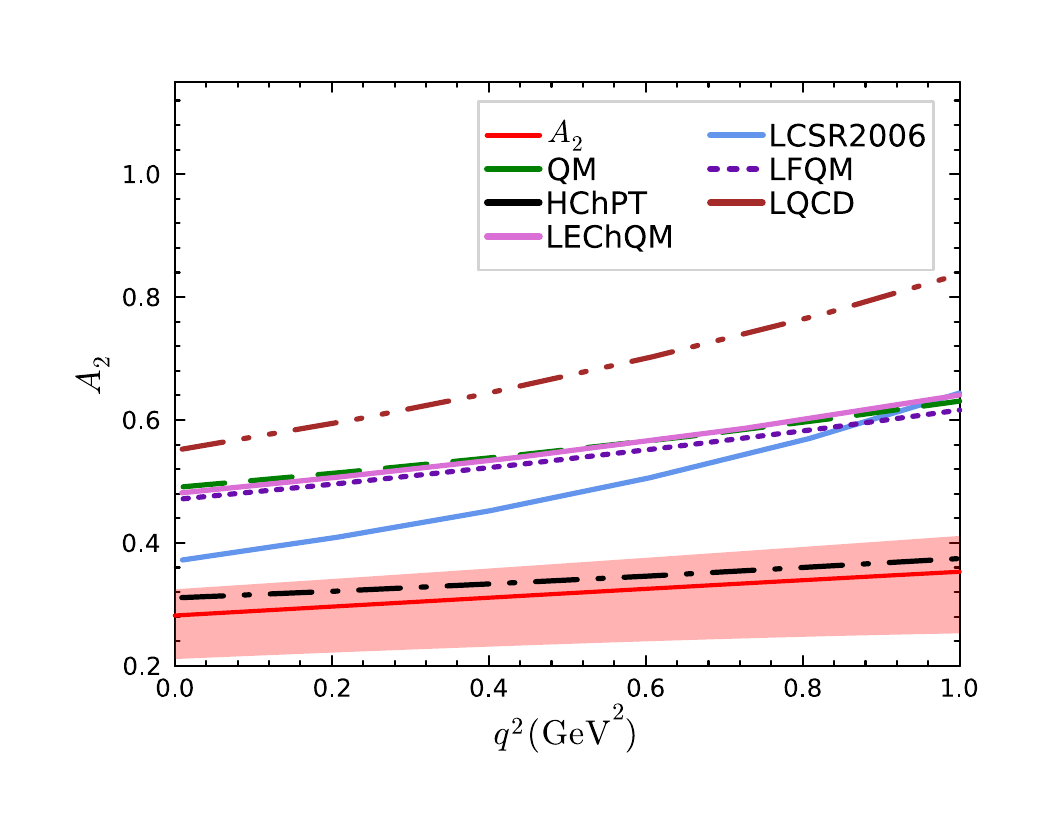}
\includegraphics[width=0.465\linewidth]{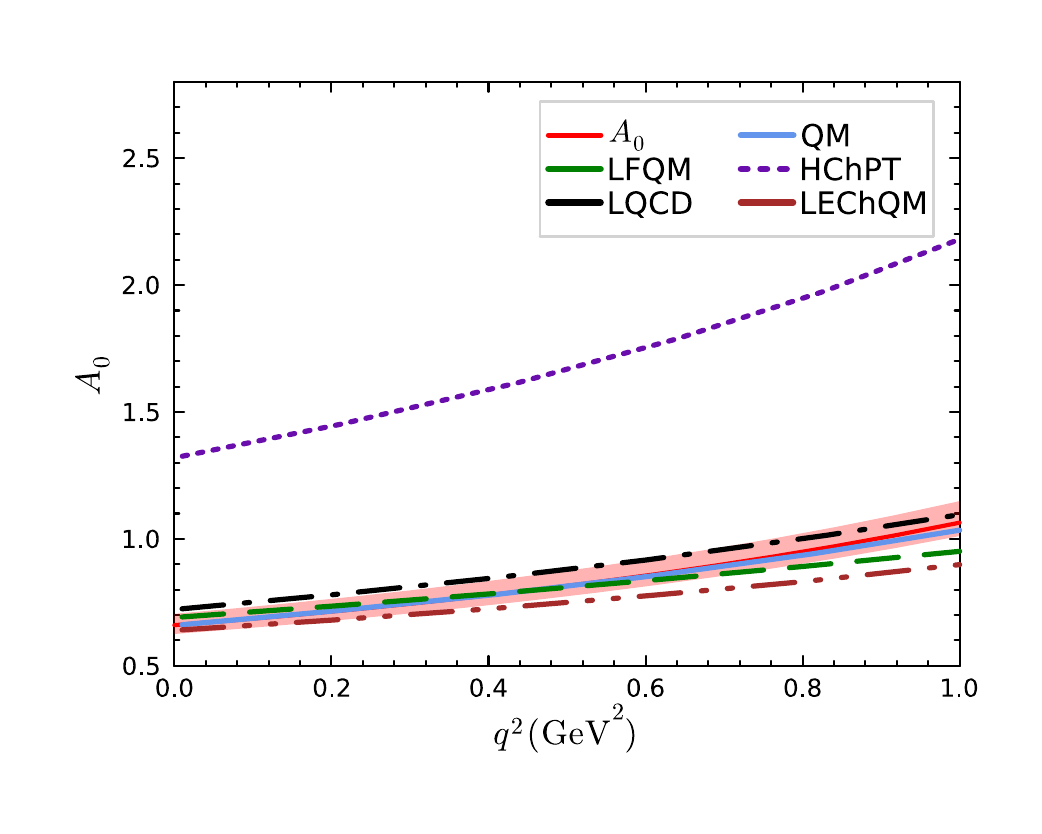}\vspace{-6mm}
\caption{LCSR predictions of the $D^0 \to \rho^-$ form factors in the whole kinematical region. Other determinations using LCSR in 2006~\cite{Wu:2006rd}, QM~\cite{Melikhov:2000yu}, LFQM~\cite{Verma:2011yw}, LEChQM~\cite{Palmer:2013yia}, HChPT~\cite{Fajfer:2005ug} and LQCD~\cite{Abada:2002ie} are shown for easy comparison.}
\label{fig:D+2rho01-whole}
\end{figure}
%---------------------------------------------------------

\begin{figure}[t]\vspace{-4mm}
\centering \includegraphics[width=0.465\linewidth]{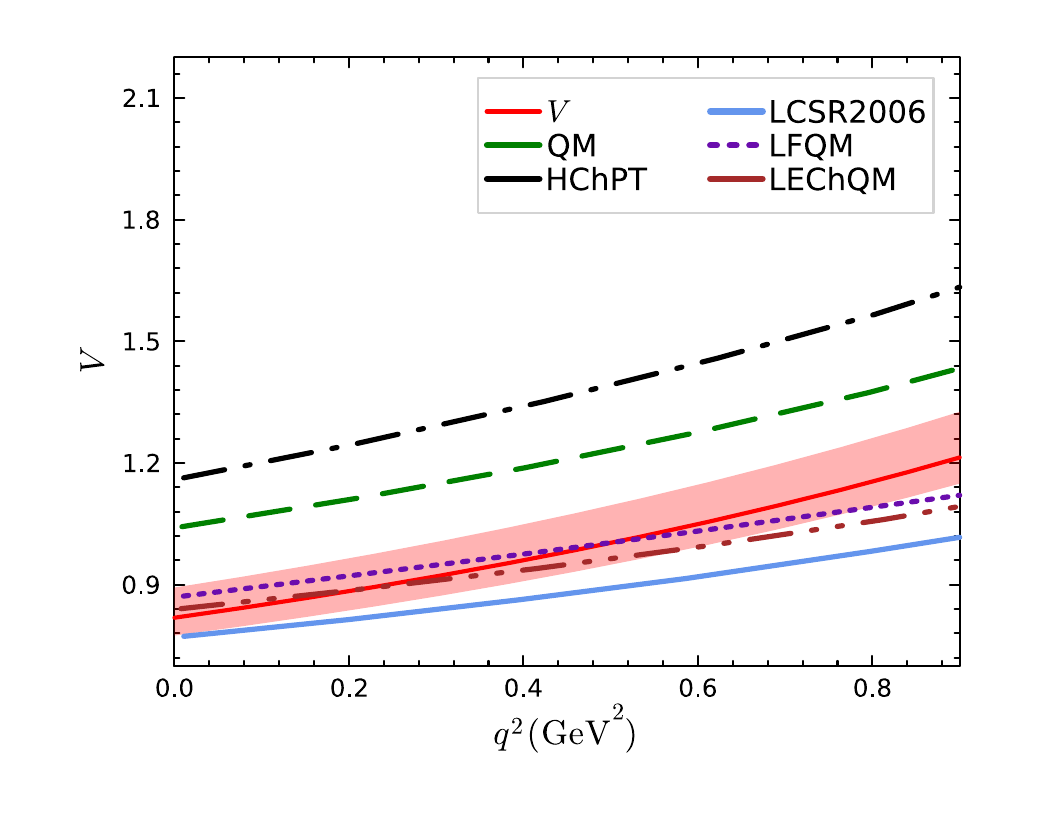}
\includegraphics[width=0.465\linewidth]{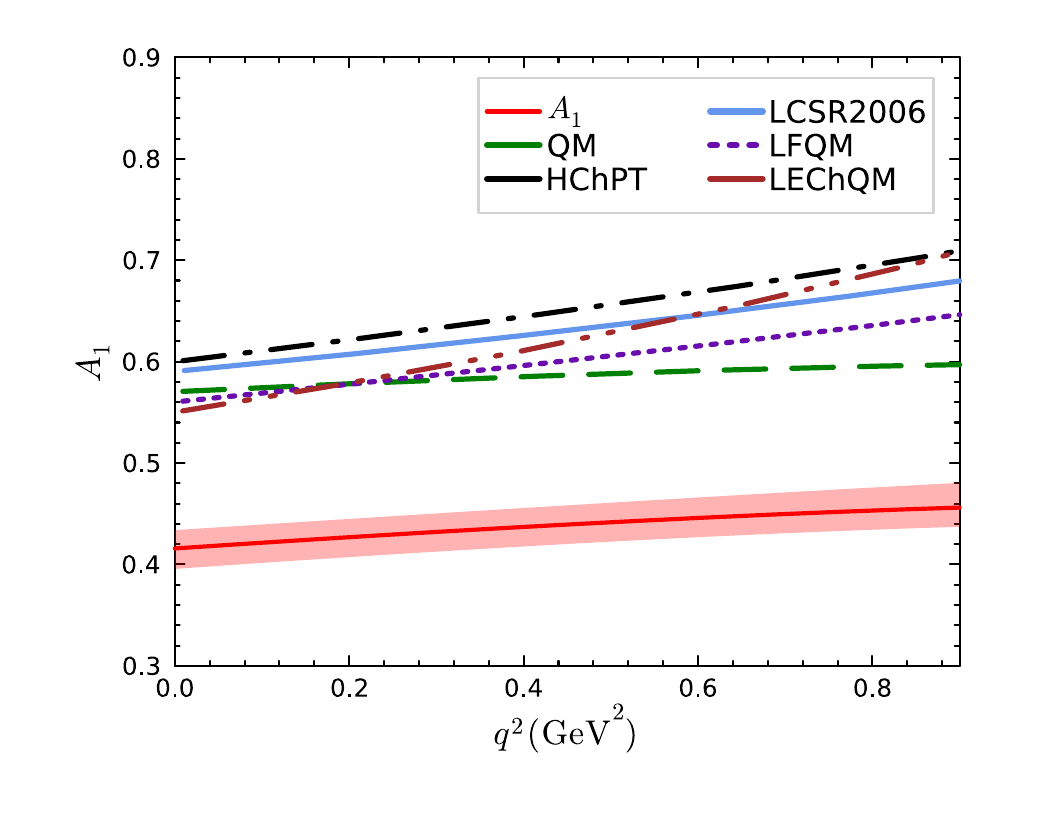}\non\vspace{-4mm} 
\includegraphics[width=0.465\linewidth]{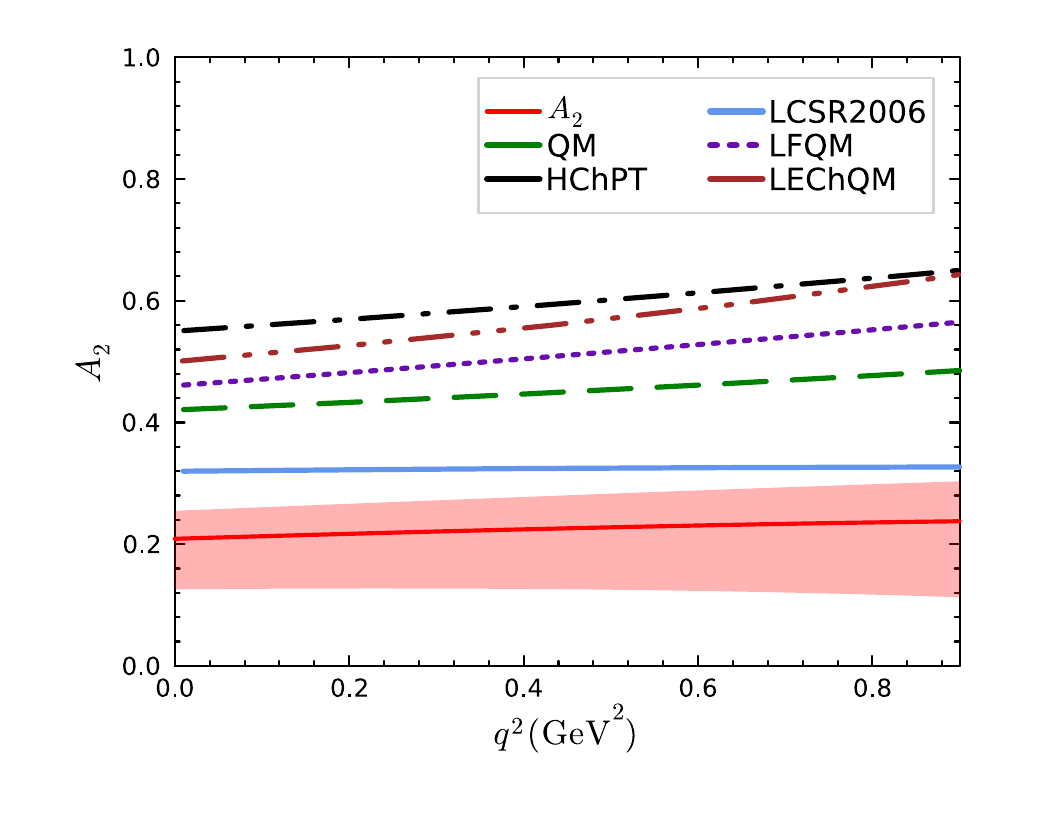}
\includegraphics[width=0.465\linewidth]{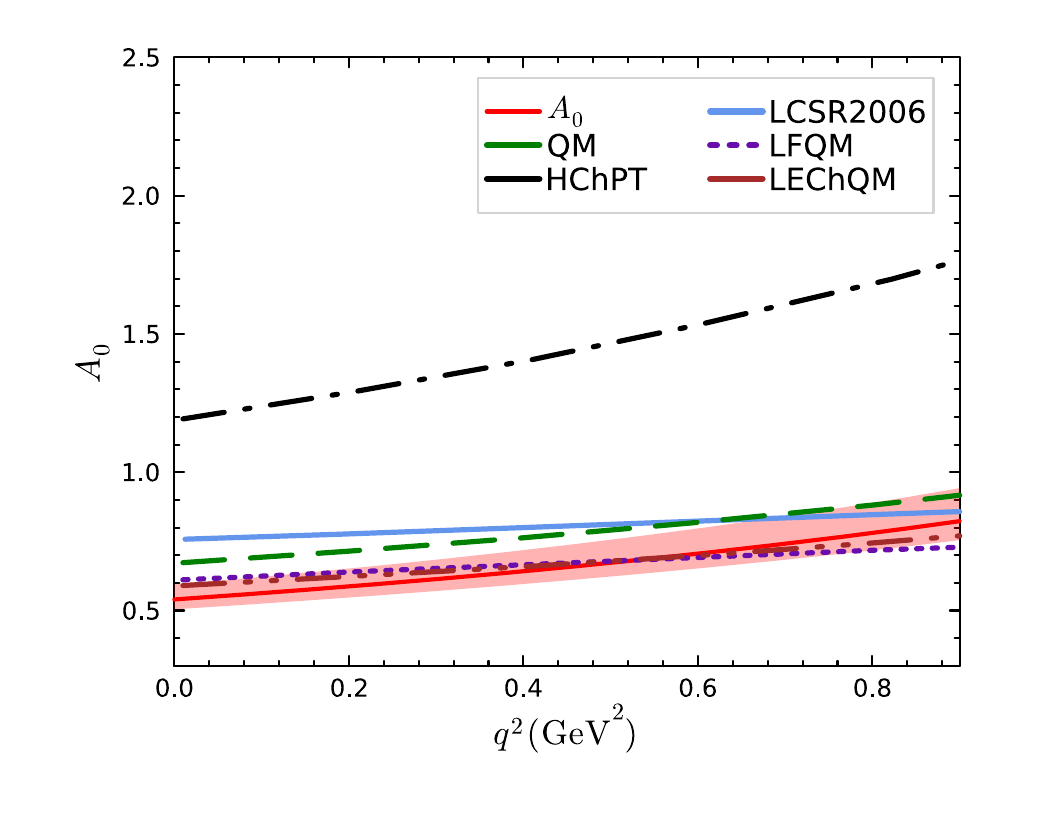}
\vspace{-6mm}
\caption{LCSR predictions of $D_s^+ \to K^{\ast 0}(892)$ form factors in the whole kinematical region.} 
\label{fig:Ds+2Ks01-whole}
\end{figure}

\begin{figure}[t]\vspace{-4mm}
\centering \includegraphics[width=0.465\linewidth]{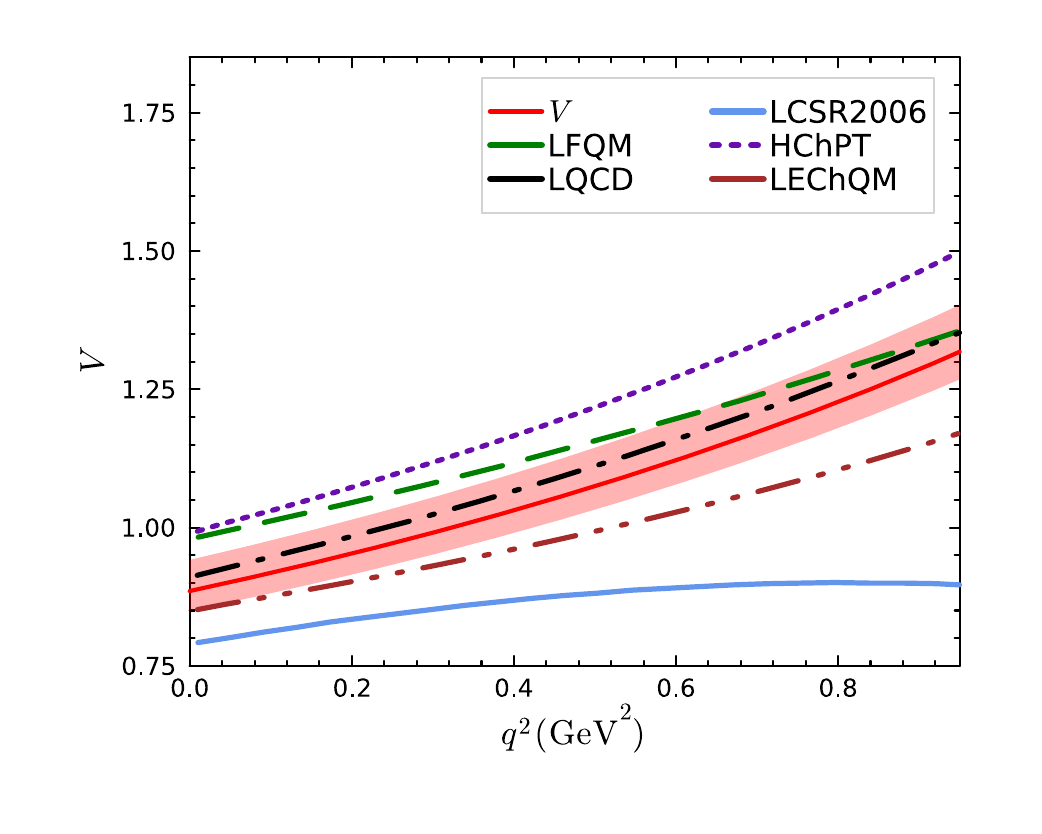}
\includegraphics[width=0.465\linewidth]{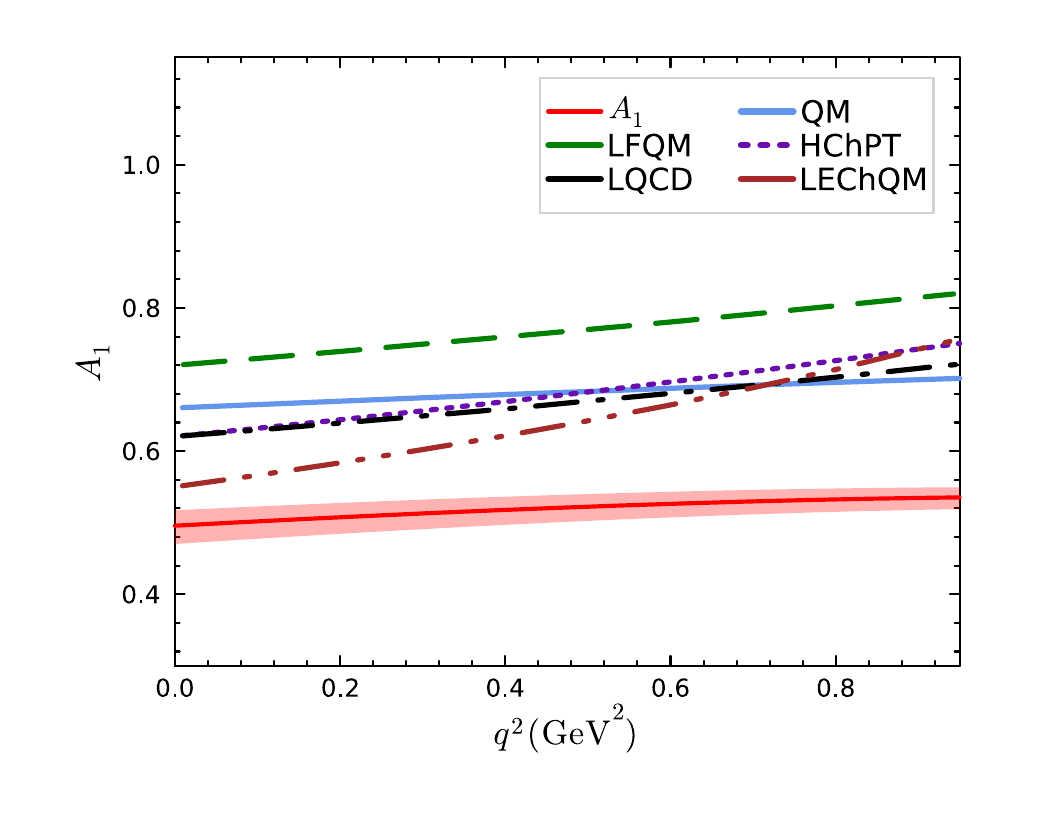}\non\vspace{-4mm}
\includegraphics[width=0.465\linewidth]{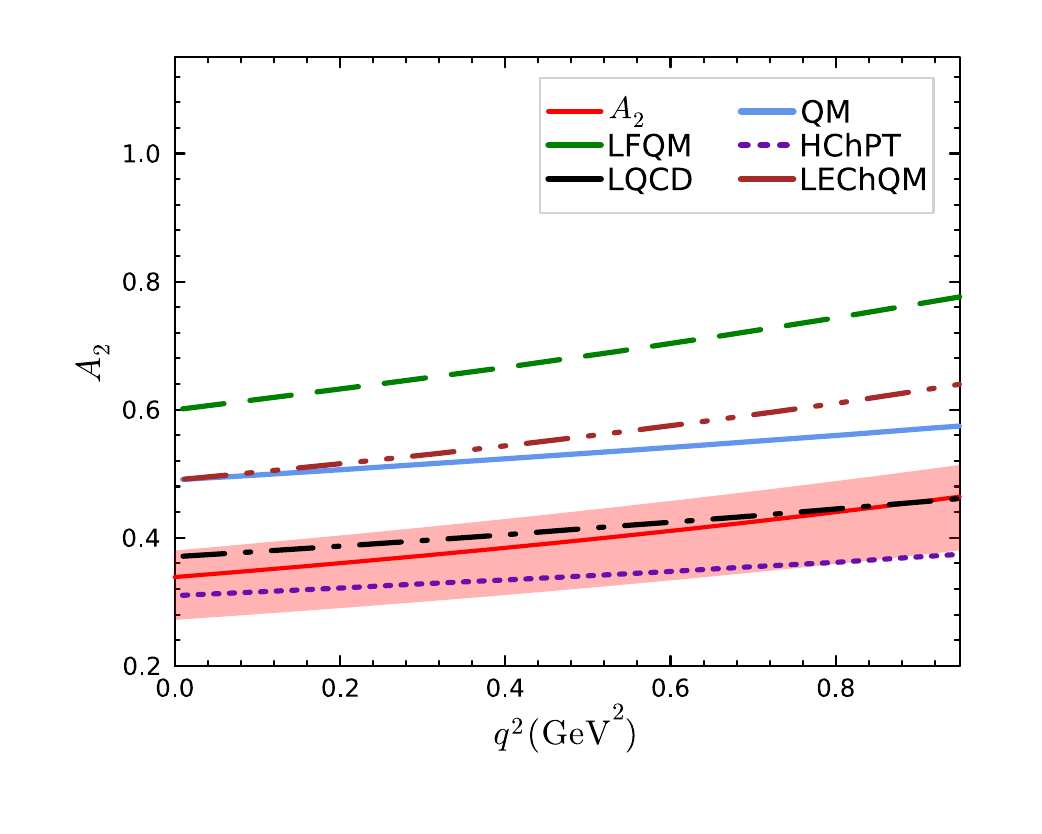}
\includegraphics[width=0.465\linewidth]{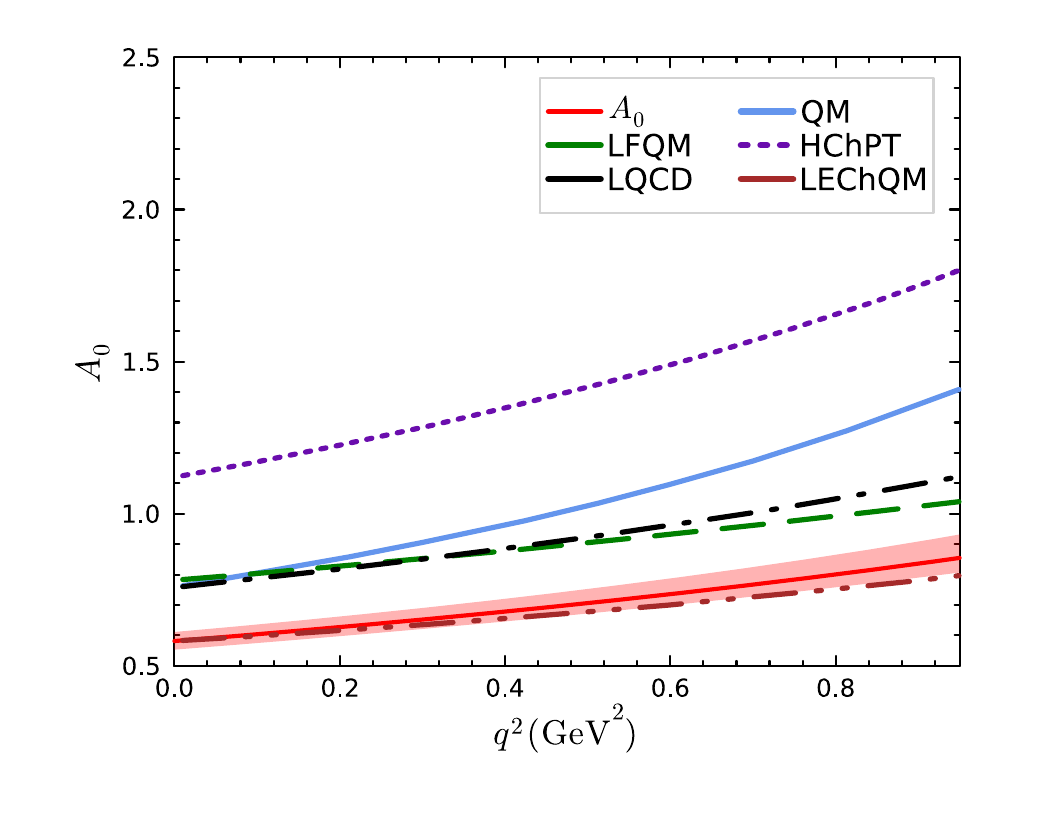}
\vspace{-6mm}
\caption{LCSR predictions of $D^+ \to K^{\ast 0}(892)$ form factors in the whole kinematical region.}
\label{fig:D+2Ks01-whole}
\end{figure}

To proceed, we plot the hadronic FFs in the whole kinematical regions for $D\to \rho$, $D_s\to K^\ast$ and $D\to K^\ast$ in Figs.~\ref{fig:D+2rho01-whole},~\ref{fig:Ds+2Ks01-whole} and~\ref{fig:D+2Ks01-whole}, respectively. Our LCSR predictions, extrapolating to small recoils using BCL parametrization, are compared with the results obtained by other theoretical approaches. 
From Figs.~\ref{fig:D+2rho01-whole},~\ref{fig:Ds+2Ks01-whole} and~\ref{fig:D+2Ks01-whole}, we find that our results of the FFs $V,A_0$ are consistent with other theoretical approaches within uncertainties, showing particularly good agreement with various quark model predictions. However, our updated LCSR predictions for $A_1$ and $A_2$ exhibit significantly lower values compared to previous calculations~\cite{Wu:2006rd}. In comparison with the earlier LCSRs result~\cite{Wu:2006rd}, our improved calculation highlights substantial power corrections at ${\cal O}(1/m_c)$ in charm quark decays, underscoring their importance in these form factors. 

%---------------------------------------------------------
%---------------------------------------------------------
%---------------------------------------------------------

\subsection{Predictions of differential decay rates}\label{semileptonicdecays}

The differential decay width for semileptonic charm decays can be expressed in terms of helicity amplitudes as a function of the momentum transfer squared~\cite{HFLAV:2022esi}, i.e.,
\beq \frac{d\Gamma\left(D_{(s)} \rightarrow Vl^{+}\nu_{l}\right)}{dq^2} = \frac{G_F^2 |V_{cq^\prime}|^2 | \vec{p} | q^2 v^2}{12(2\pi)^3 m_{D_{(s)}}^2} \Big[ \left( 1 - \delta_l\right) \sum_{i=+,-,0} \vert H_i(q^2) \vert^2 - 3 \delta_l \vert H_t(q^2) \vert^2 \Big] \ , \eeq
with $\vert \vec{p} \vert$ representing the magnitude of three-momentum of the vector meson. 
Here, $G_F =1.17 \times 10^{-5} $ GeV$^2$ is the Fermi constant, and $V_{cq^\prime}$ ($q^\prime=d,s$) denotes the CKM matrix element. In our case, we adopt the world average values from the PDG \cite{ParticleDataGroup:2024cfk}: $\vert V_{cd} \vert =0.225, \vert V_{cs} \vert =0.973$. Further more, $\delta_l = m_l^2/(2q^2)$ characterizes the helicity-flip effect and the velocity parameter is given by $v=1-m_l^2/q^2$. Tthe helicity amplitudes are associated with the FFs in Eq.~\eqref{eq:FFs-def1} through
\beq 
&&H_t(q^2) = -\frac{2\mathrm{i}m_D |\vec{p}_2|}{\sqrt{q^2}} A_0(q^2), \qquad H_{\pm}(q^2) = -(m_D + m_V)\mathrm{i} A_1(q^2) \pm \frac{2\mathrm{i}m_D |\vec{p}|}{m_D + m_V} 2V(q^2), \non
&&H_0(q^2) = -\frac{\mathrm{i}(m_D + m_V)}{2m_V\sqrt{q^2}} \left(m_D^2 - m_V^2 - q^2\right) A_1(q^2) + \frac{\mathrm{i}}{m_D + m_V} \frac{2m_D^2 |\vec{p}|^2}{m_V\sqrt{q^2}} A_2(q^2).\label{eq.hel.amp}
\eeq
The longitudinal ($H_0$) and transversal form factors ($H_{\pm}$) are both relevant to axial-vector form factor $A_1$. The form factor $H_t$, corresponding to the timelike-helicity of the dilepton, does not contribute to differential decay width in the limit of zero lepton mass. 

\begin{figure*}[t]\vspace{-4mm} \centering 
\includegraphics[width=0.4\linewidth,height=5.6cm]{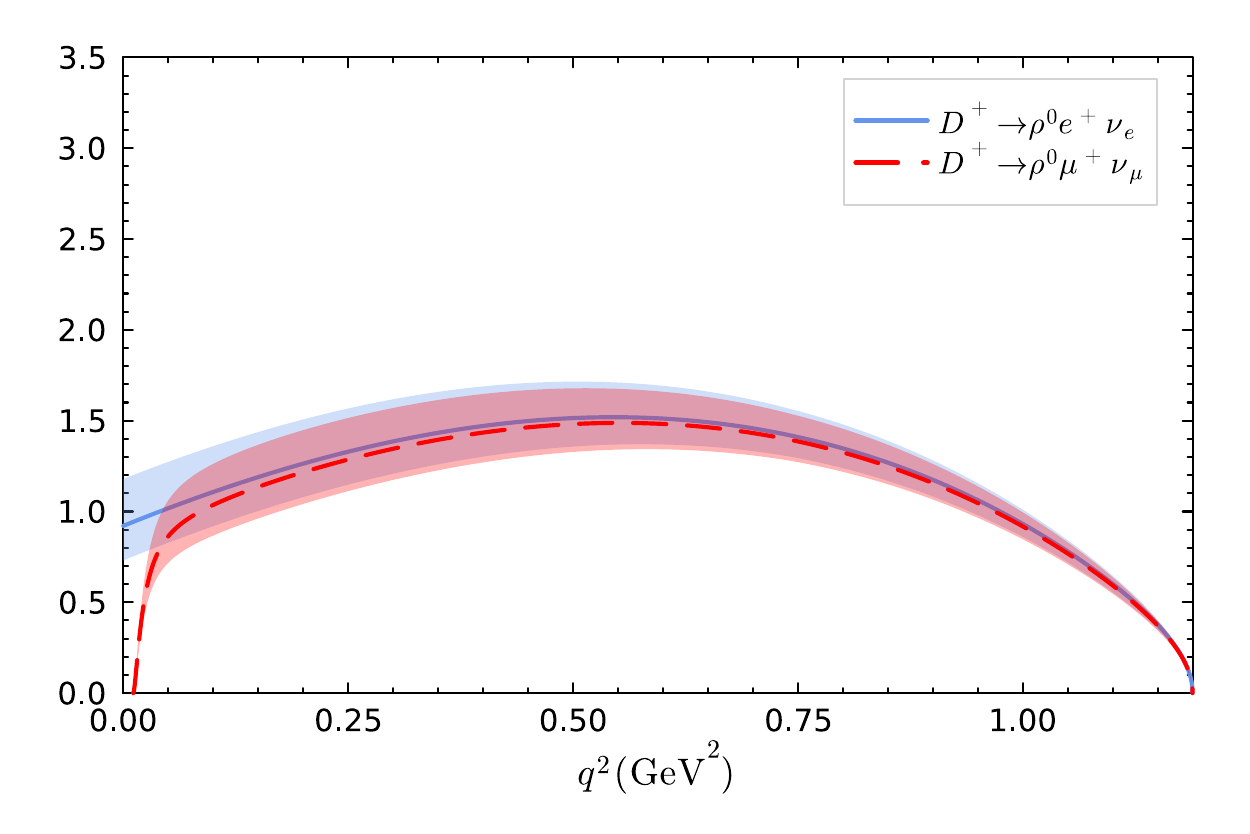}\hspace{4mm}
\includegraphics[width=0.4\linewidth,height=5.6cm]{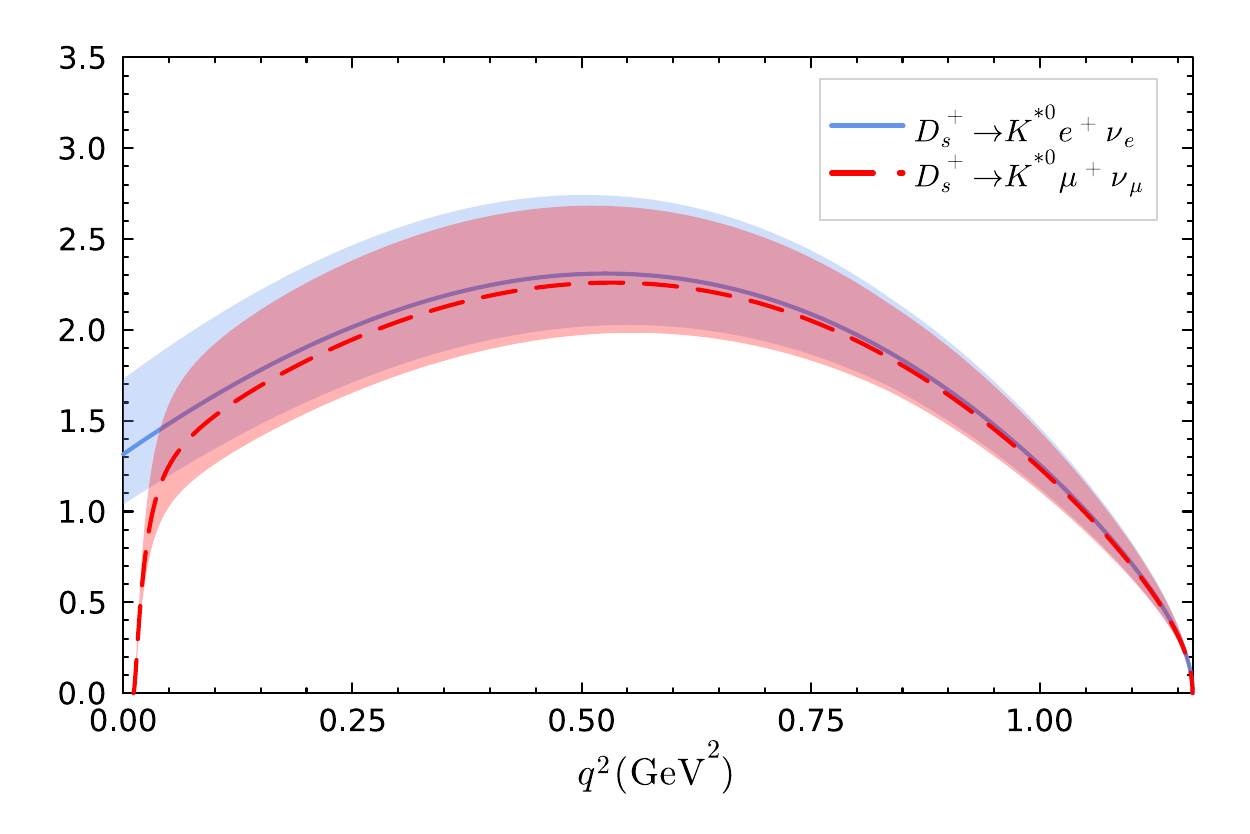} \non\vspace{-2mm}
\includegraphics[width=0.4\linewidth,height=5.6cm]{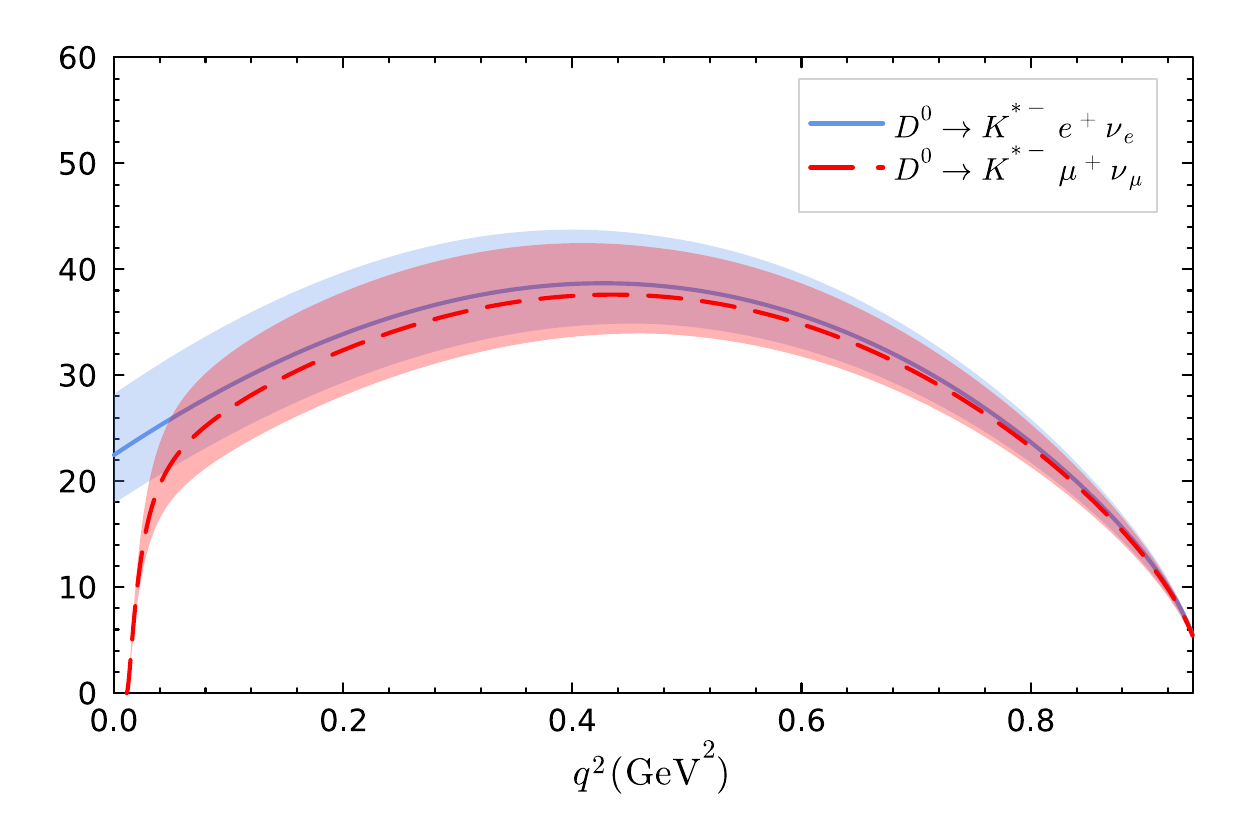}\hspace{4mm}
\includegraphics[width=0.4\linewidth,height=5.6cm]{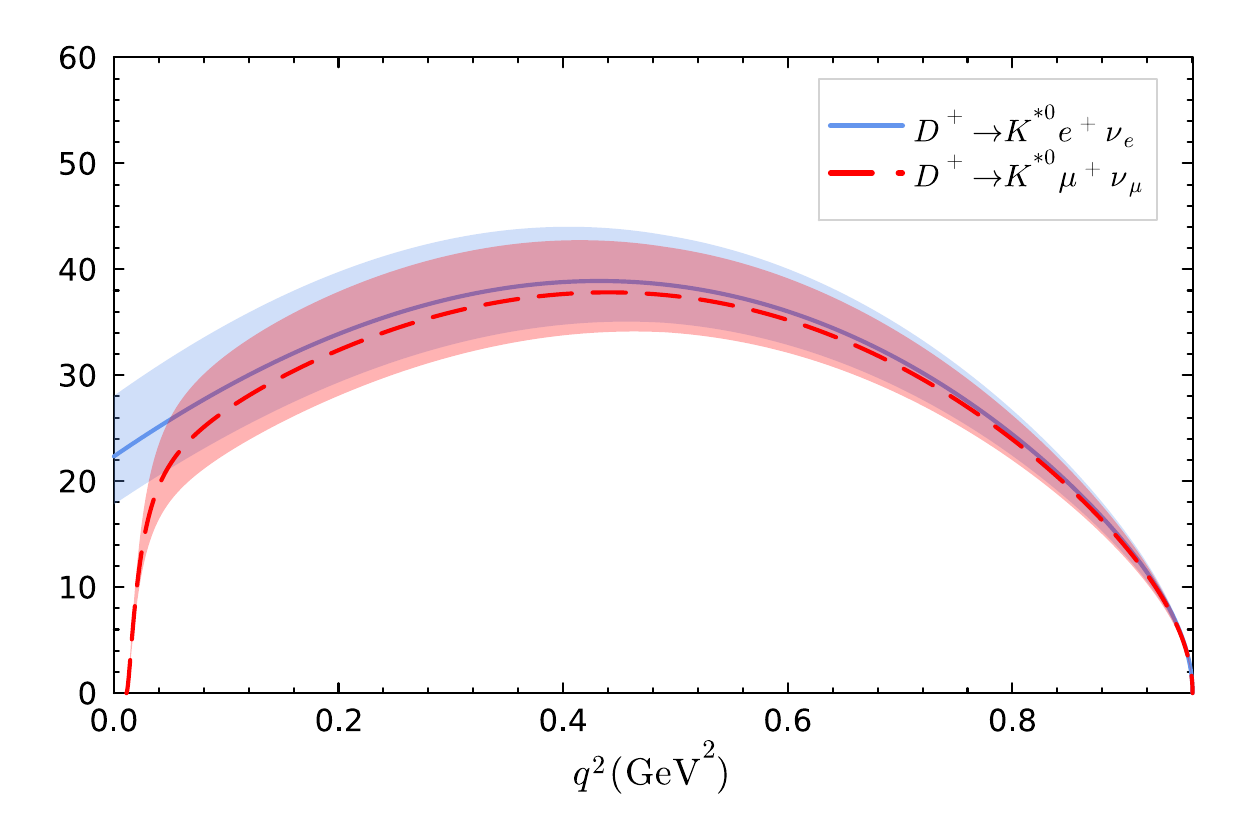} 
\vspace{-4mm}
\caption{Differential decay widths $d\Gamma/dq^2$ (in unit of $10^{-15}$ GeV$^{-1}$) 
induced by the $c \to d$ (up) and $c \to s$ (down) weak currents. }
\label{fig:c2dlnu-c2slnu}
\end{figure*}

Based on the results obtained in the previous section, the differential decay widths for semileptonic charm decays, induced by $c \to d$ and $c \to s$ weak currents, are displayed in Figure \ref{fig:c2dlnu-c2slnu}. It can be seen from the figures that the lepton mass effect is mainly pronounced in the large recoil region, where the momentum transfer approaches zero. At intermediate and large momentum transfers, however, the mass effect is slight and can be safely neglected. This observation is consistent with the finding of the CLFQM calculation \cite{Zhang:2020dla}.

%---------------------------------------------------------
%---------------------------------------------------------

In Table \ref{tab:D2V_BR}, we present the LCSR predictions for the branching ratios of the four $D_{(s)} \to V l^+\nu_l$ decays under our consideration. Note that the mean lifes $\tau_{D^+} = 1040 \times 10^{-15} s$, $\tau_{D^0} = 410 \times 10^{-15} s$ and $\tau_{D_s^+} = 504 \times 10^{-15} s$ have been used to obtain the branching ratios. For comparison, we also show results from previous LCSR calculation~\cite{Wu:2006rd}, the covariant light-front quark model (CLFQM)~\cite{Zhang:2020dla}, the HChPT \cite{Fajfer:2005ug} and the experimental values from PDG \cite{ParticleDataGroup:2024cfk}. In comparison to the previous LCSR calculation~\cite{Wu:2006rd}, 
the corrections arising from higher-twist LCDAs and higher-order terms in the lower-twist LCDA expansion are incorporated in this work, which turn out to play a significant role in the $K^\ast$ channels. This implies a potentially sizable SU(3) flavor symmetry breaking effect. Our results are marginally consistent with the world average values~\cite{ParticleDataGroup:2024cfk}, given that the uncertainties are taken into account. 
At the current accuracy, there still exists a $10\%-20\%$ discrepancy when confronting our LCSR predictions with experimental measurements. This difference indicates the necessity to further implement the effects of vector meson widths and non-resonant QCD background in the semileptonic charmed meson decays to vector mesons. Our findings appeal for dedicated studies of four-body semileptonic charm decays $D_{(s)} \to K\pi l \nu$, from both theoretical \cite{D2pipi-LCSRs} and experimental \cite{Belle:2020xgu} perspective.

%---------------------------------------------------------
\begin{table*}[tb]
\caption{Branching ratios of the semileptonic charm decays $D_{(s)} \to V l^+\nu_l$ (in unit of $10^{-3}$).}
\label{tab:D2V_BR}
\centering 
\renewcommand{\arraystretch}{1.2} % 
\setlength{\tabcolsep}{6pt}       % 
\begin{tabular}{c|cccc}
\toprule\toprule
& 
\makecell{\textbf{$D^+ \to \rho^0 \ell^+\nu_\ell$}} & 
\makecell{\textbf{$D_s^+ \to K^{\ast0} \ell^+\nu_\ell$}} & 
\makecell{\textbf{$D^0 \to K^{\ast-} \ell^+\nu_\ell$}} & 
\makecell{\textbf{$D^+ \to \bar{K}^{\ast0} \ell^+\nu_\ell$}}  \\
\midrule
This work & \makecell{$2.30^{+0.32}_{-0.25}$ \\ $2.20^{+0.30}_{-0.23}$  } & 
\makecell{$1.55^{+0.30}_{-0.20}$ \\ $1.48^{+0.29}_{-0.19}$  } & \makecell{$17.6^{+2.4}_{-1.9}$ \\ $16.6^{+2.2}_{-1.8}$  } & \makecell{$45.2^{+6.2}_{-5.0}$  \\ $42.7^{+5.7}_{-4.5}$  } \\ \addlinespace
LCSR2006 \cite{Wu:2006rd} & 
\makecell{$2.29^{+0.23}_{-0.16}$  \\ $2.20^{+0.21}_{-0.16}$  } & \makecell{$2.33^{+0.29}_{-0.30}$  \\ $2.24^{+0.27}_{-0.29}$  } & \makecell{$21.2\pm0.9$ \\ $20.1 \pm 0.9$  } & \makecell{$53.7^{+2.4}_{-2.3}$  \\ $51.0^{+2.3}_{-2.1}$  } \\  \addlinespace
CLFQM \cite{Zhang:2020dla} & \makecell{2.32  \\ 2.22  } & 
\makecell{1.90  \\ 1.82  } & --- & 
\makecell{73.2  \\ 69.3  } \\\addlinespace
HChPT \cite{Fajfer:2005ug} & 2.50 & 2.20 & 22.0 & 56.0 \\\addlinespace
PDG \cite{ParticleDataGroup:2024cfk} & 
\makecell{$1.90\pm0.10$  \\ $2.40 \pm 0.40$  } & 
\makecell{$2.15 \pm 0.28$ \\ ---} & \makecell{$21.5 \pm 1.60$ \\ $18.9 \pm 2.40$  } & \makecell{$54.0 \pm 1.00$  \\ $52.7 \pm 1.50$  } \\
\bottomrule\bottomrule
\end{tabular}
\end{table*}
%---------------------------------------------------------

\section{Summary and Outlook}\label{sec:summary}

Motivated by recent experimental progress in semileptonic charm decays and advances in the LCDAs description of vector mesons, we investigate $D_{(s)} \to V l \nu_l$ ($V=\rho,K^{\ast}$) using the QCD LCSRs approach. 
Compared to previous studies, our analysis incorporates: higher-twist corrections from both two- and three-particle LCDAs, higher-power QCD corrections to the lower-twist two-particle LCDAs, and power correction in the heavy quark expansion at the order of ${\cal O}(1/m_c)$. Our calculation demonstrates that the LCSR framework maintains robust predictive power for the $D_{(s)} \to V$ transitions, despite the inherent challenges regarding the intermediate charm quark mass scale and the significant width effects of the final vector mesons.

Our main findings are summarized as follows. 
Firstly, both the three-particle LCDAs and the power correction from the heavy quark expansion at ${\cal O}(1/m_c)$ 
contribute significantly to the form factors $A_1$ and $A_2$. 
Secondly, The OPE expansion exhibits good convergence. While two-particle twist-three LCDAs contribute sizeably or even dominantly in the axial-vector transition, the higher-twist effects remain well under control. Last but not least, the lepton mass effect is mostly prominent in the large-recoil region of the semileptonic charm to vector meson decays. Additionally, a sizable SU(3) flavor-breaking effect is observed in $D_{(s)} \to K^\ast l \nu$ decays.  

Follow-up studies along this work will focus on refining the $D_{(s)} \to V$ transition form factors. A comprehensive study  can be conducted across the full momentum transfer range,  incorporating inputs from other theoretical approaches and lattice QCD to reduce model dependence in the BCL parameterization when extrapolating to intermediate and large momentum transfers. Calculation of next-to-leading-order QCD corrections to the form factors \cite{D2VNLO} will help constrain the inherent uncertainty associated with the choice of the charm quark mass. Furthermore, a direct investigation of the finite resonant width effect and non-resonant QCD backgrounds in four-body semileptonic decays via di-meson LCDAs and dispersive techniques~\cite{Yao:2020bxx} will also be promising to improve the LCSR predictions obtained in the current work.

\acknowledgments
We are grateful to Hai-Long Ma, Yu-Ming Wang, Shu-Lei Zhang and Ming Zhong for fruitful discussions. 
This work is supported by the National Key R$\&$D Program of China under Contracts No. 2023YFA1606000; by the National Science Foundation of China (NSFC) under Contract No.~12275076, No.~12335002 and No. 11975112; by the Science Fund for Distinguished Young Scholars of Hunan Province under Grant No.~2024JJ2007; by the Fundamental Research Funds for the Central Universities under Contract No.~531118010379.

\appendix
\section{LCDAs of light vector mesons}\label{app:LCDAs}

In order to facilitate the light-cone expansion, the meson four-momentum $p_\mu$ and the near lightlike separation $x_\mu$ are expressed as linear combinations of the lightlike vectors ${\hat p}_\mu$ and $z_\mu$~\cite{Chernyak:1983ej}, 
\beq  {\hat p}_\mu = p_\mu - \frac{1}{2} z_\mu \frac{m_V^2}{{\hat p} \cdot z}, \qquad 
z_\mu = x_\mu \left[ 1 - \frac{x^2 m_V^2}{4({\hat p} \cdot z)^2} \right] - \frac{1}{2} {\hat p}_\mu \frac{x^2}{{\hat p} \cdot z} + \mathcal{O}(x^4)\,.
\label{eq:lc-expansion}\eeq
Meanwhile, the polarization vector is decomposed as, 
\beq \eta^\ast_\mu = \frac{\eta^\ast \cdot x}{{\hat p} \cdot z} \, {\hat p}_\mu + \frac{\eta^\ast \cdot {\hat p}}{{\hat p} \cdot z }\, z_\mu + \eta^\ast_{\perp \mu} 
= (\eta^\ast \cdot x) \, \frac{p_\mu (p \cdot x) - x_\mu m_V^2}{(p \cdot x)^2 - x^2 m_V^2}\ . \label{eq:pola} \eeq

\subsection{Two-particle LCDAs}

LCDAs are rigorously defined by the matrix elements sandwiched by the quark bilinears with light-cone separation, 
and are switched to the actual momenta and the near lightlike distance $x$ in practice. In Refs.~\cite{Ball:1998sk,Ball:1998ff}, high-twist LCDAs of vector mesons are systematical studied in QCD 
based on conformal expansion with meson and quark mass corrections taken into account. The complete analysis of the parameters from QCD sum rules and a renormalon based model is presented in Refs.~\cite{Ball:2007rt,Ball:2007zt}. 
With the polarisation decomposition in Eq.~\eqref{eq:pola}, 
the matrix elements sandwiched between vacuum and vector meson state read
\beq
&&\langle V(p,\eta^\ast) \vert {\bar q_1}(0) \gamma_\mu q_2(x) \vert 0 \rangle 
= f_V^\parallel m_V \int_0^1 du e^{i {\bar u} p \cdot x} \Big\{ \eta^\ast_\mu \left[ \phi_3^\perp(u) + \frac{m_V^2 x^2}{16} \phi_5^\perp(u) \right] \non
&& \hspace{4.2cm} + p_\mu \frac{\eta^\ast \cdot x}{p \cdot x} \left[ \phi_2^\parallel(u) - \phi_3^\perp(u) + \frac{m_V^2 x^2}{16}  \left( \phi_4^\parallel(u) - \phi_5^\perp(u) \right) \right] \non
&& \hspace{4.2cm} - \frac{(\eta^\ast \cdot x) m_V^2}{2 (p \cdot x)^2} x_\mu \Big[ \psi_4^\parallel(u) - 2 \phi_3^\perp(u) + \phi_2^\parallel(u) \Big] \Big\}\ , 
\label{eq:LCDAs-definition-1}\\
&&\langle V(p,\eta^\ast) \vert {\bar q_1}(0) \sigma_{\mu\nu} q_2(x) \vert 0 \rangle 
= - i f^\perp_V \int_0^1 du e^{i {\bar u} p \cdot x} \Big\{ 
\left( \eta^\ast_\mu p_\nu - \eta^\ast_\nu p_\mu \right) \left[ \phi_2^\perp(u) + \frac{m_V^2 x^2}{16} \phi_4^\perp(u) \right] \non
&& \hspace{4.2cm} + \left( p_\mu x_\nu - p_\nu x_\mu \right) \frac{(\eta^\ast \cdot x) m_V^2}{(p \cdot x)^2} 
\left[ \phi_3^\parallel(u) - \frac{1}{2} \phi_2^\perp(u) - \frac{1}{2} \psi_4^\perp(u)\right]  \non
&& \hspace{4.2cm} + \frac{m_V^2}{2(p \cdot x)}\left( \eta^\ast_\mu x_\nu - \eta^\ast_\nu x_\mu \right) \Big[ \psi_4^\perp(u) - \phi_2^\perp(u) \Big] \Big\}\ , 
\label{eq:LCDAs-definition-2}\\
&&\langle V(p,\eta^\ast) \vert {\bar q_1}(0) \gamma_\mu \gamma_5 q_2(x) \vert 0 \rangle 
=  \frac{ f_V^\parallel m_V \varepsilon_{\mu\nu\rho\sigma} \eta^{\ast \nu} p^\rho x^\sigma}{4} \int_0^1du e^{i {\bar u} p \cdot x} 
\left[ \tilde{\psi}_3^\perp(u) + \frac{m_V^2 x^2}{16} \tilde{\psi}_5^\perp(u) \right],
\label{eq:LCDAs-definition-2}\\
&&\langle V(p,\eta^\ast) \vert {\bar q_1}(0) q_2(x) \vert 0 \rangle 
=  \frac{i}{2} f_V^\perp (\eta^\ast \cdot x) m_V^2 \int_0^1du e^{i {\bar u} p \cdot x} \tilde{\psi}_3^\parallel(u) \ .
\eeq
We use the conventions 
\beq \varepsilon_{0123} = - \varepsilon^{0123} = 1, \quad
\gamma_5 = \frac{i}{4!} \varepsilon^{\mu\nu\rho\sigma} \gamma_\mu \gamma_\nu \gamma_\rho \gamma_\sigma. \eeq
The quark mass effects are taken into account by the matrix elements with the Dirac structures $\gamma_\mu \gamma_5$ and ${\bf 1}$. The following auxillary DAs have been introduced,  
\beq \tilde{\psi}_3^\parallel(u) = \left( 1 - r_\parallel \delta_+ \right) \psi_3^\parallel(u)\ , \qquad 
\tilde{\psi}_{3(5)}^\perp(u) = \left( 1 - r_\perp \delta_+ \right) \psi_{3(5)}^\perp(u)\ , \eeq
with $r_\parallel = f_V^\parallel/f_V^\perp$, $r_\perp = f_V^\perp/f_V^\parallel$ and $\delta_\pm = \left( m_{q_2} \pm m_{q_1} \right)/m_V$. 
The DAs $\phi = \{ \phi_2^{\parallel(\perp)}, \phi_3^{\parallel(\perp)}, \psi_3^{\parallel(\perp)}, \psi_4^{\parallel(\perp)} \}$ 
satisfy the normalizations
\beq \int_0^u \phi(u^\prime) du^\prime \Big\vert_{u=1} = 1, \qquad  \int_0^u du^\prime \int_0^{u^\prime} \phi(u^{\prime\prime}) du^{\prime\prime} \Big\vert_{u=1} = 1 \ ,
\eeq 
and also the equation of motions (EOM) of the LCDAs. 
The DAs $\phi^\prime = \{ \phi_4^{\parallel(\perp)}, \phi_5^\perp, \psi_5^\perp \}$ 
are not subject to a particular normalisation, nevertheless, $\int_0^u du^\prime \left( \phi_4^\parallel - \phi_5^\perp \right) \big\vert_{u=1} = 0 $ is necessary. They are related to the DAs $\phi$, at first order of ${\cal O}(m_V^2)$ expansion, by~\cite{Ball:1998ff}
\beq &&\phi_4^\parallel(u) = - 4 \int_0^u \left[ (2 u^\prime -1) \phi_2^\parallel(u^\prime) \right] 
+ 4 \int_0^u du^\prime \int_0^{u^\prime} du^{\prime\prime} 
\left[ \phi_3^\perp(u^{\prime\prime}) - \psi_4^\parallel(u^{\prime\prime}) - 3 \phi_2^\parallel(u^{\prime\prime}) \right],  \nonumber\\
&&\phi_4^\perp(u) = - 4 \int_0^u \left[ (2 u^\prime -1) \phi_2^\perp(u^\prime) \right] 
+ 4 \int_0^u du^\prime \int_0^{u^\prime} du^{\prime\prime} \left[ \psi_4^\perp(u^{\prime\prime}) - \phi_2^\perp(u^{\prime\prime}) \right],  \nonumber\\
&&\phi_5^\perp(u) = - 4 \int_0^u \left[ (2 u^\prime -1) \phi_3^\perp(u^\prime) \right], \nonumber\\ 
&&\psi_5^\perp(u) = - 4 \int_0^u \left[ (2 u^\prime -1) \psi_3^\perp(u^\prime) \right].
\label{eq:phi-phip-relations} \eeq
We note that the last two relations hold only for asymptotic LCDAs~\cite{Bharucha:2015bzk}.

The low-twist DAs are conventionally expanded in conformal spin, which is analogous to the partial wave expansion of ${\rm SO(3)}$, 
and written in terms of Gegenbauer polynomials with corresponding moments. 
In this work, we take the truncation at the second order of leading twist DAs expansion, 
\beq \phi_2^{\parallel(\perp)}(u) = 6u(1-u) \left[ 1 + a_2^{\parallel(\perp)} C_2^{3/2}(t) \right]. \label{eq:phi2} \eeq
The twist-3 DAs contributed from the leading twist DAs are cited as \cite{Ball:1998sk}
\beq &&\tilde{\psi}_3^\parallel(u) = {\bar u} \int_0^u du^\prime \frac{\Psi^\parallel_2(u^\prime)}{{\bar u}^\prime} 
+ u \int_u^1 du^\prime \frac{\Psi^\parallel_2(u^\prime)}{u^\prime} + 6 u {\bar u} \left[ \frac{5}{18} \omega_3^\perp C_2^{3/2}(2u-1) \right], 
\label{eq:psi3para} \nonumber\\
&&\phi_3^\parallel(u) = \frac{1}{2} \int_0^u du^\prime \frac{\Psi^\parallel_2(u^\prime)}{{\bar u}^\prime} 
+ \frac{1}{2} \int_u^1 du^\prime \frac{\Psi^\parallel_2(u^\prime)}{u^\prime} + \frac{15}{8} \omega_3^\perp \left[ 3 - 30(2u-1)^2 + 35 (2u-1)^4 \right], 
\label{eq:phi3para} \nonumber\\
&&\tilde{\psi}_3^\perp(u) = {\bar u} \int_0^u du^\prime \frac{\Psi^\perp_2(u^\prime)}{{\bar u}^\prime} 
+ u \int_u^1 du^\prime \frac{\Psi^\perp_2(u^\prime)}{u^\prime} + 6 u {\bar u} \left[ \frac{10}{9} \zeta_3^\parallel + \frac{5}{12} 
\left( \omega_3^\parallel - \frac{\tilde{\omega}_3^\parallel}{2} \right) \right] C_2^{3/2}(2u-1), \label{eq:psi3perp} \nonumber\\
&&\phi_3^\perp(u) = \frac{1}{4} \int_0^u du^\prime \frac{\Psi^\perp_2(u^\prime)}{{\bar u}^\prime} 
+ \frac{1}{4} \int_u^1 du^\prime \frac{\Psi^\perp_2(u^\prime)}{u^\prime} + 5 \zeta_3^\parallel \left[ 3 (2u-1)^2 - 1 \right] \non
&& \hspace{1.2cm} + \frac{15}{32} \left( \omega_3^\parallel - \frac{\tilde{\omega}_3^\parallel}{2}\right) \left[ 3 - 30 (2u-1)^2 + 35 (2u-1)^4 \right], 
\label{eq:phi3perp} \eeq
with the auxiliary functions 
\beq &&\Psi^\parallel_2(u^\prime) = 2 \phi_2^\perp(u^\prime)+ r_\parallel \left[ \frac{\left(3-2 u^\prime\right)}{2} \delta_+  + \frac{\delta_-}{2} \right] 
\frac{\partial \phi_2^\perp(u^\prime)}{\partial u^\prime} \,,  \nonumber\\
&&\Psi^\perp_2(u^\prime) = 2 \phi_2^\parallel(u^\prime) + r_\perp 
\left[ \delta_+ (2 u^\prime -1) + \delta_- \right] \frac{\partial \phi_2^\perp(u^\prime)}{\partial u^\prime} \,.
\label{eq:psi2to3} \eeq

In the asymptotic limit, the twist-4 DAs, contributing at the order ${\cal O}((p \cdot x)^{-2})$, are given by 
\beq \psi_4^\parallel(u) = 6 u {\bar u} \,, \quad\, \psi_4^\perp(u) = 6 u {\bar u}, \label{eq:psi4} \eeq
the twist-4 and twist-5 DAs $\phi^\prime$, contributing at the order ${\cal O}(m_V^2x^2)$, can be read from Eq. (\ref{eq:phi-phip-relations}),
\beq &&\phi_4^\parallel(u) = 24 u^2 {\bar u}^2 \,, \quad \phi_4^\perp(u) = 24 u^2 {\bar u}^2 \,,  \label{eq:phi4} \nonumber\\
&&\phi_5^\perp(u) = 6 u {\bar u} \left( 1 - u {\bar u} \right) \,, \quad \psi_5^\perp(u) = 12 u^2 {\bar u}^2. \label{eq:phi5}\eeq

\subsection{Three-particle LCDAs}

The definition of three-particle LCDAs are quoted as \cite{Ball:1998kk}
\beq &~&\langle V(p,\eta^\ast) \vert \bar{q}_1(0) g G_{\mu\nu} (\bar{v}x) \sigma_{\alpha\beta} q_2(x)\vert0\rangle \non 
&=& f_V^\perp m_V^2 \frac{\eta^\ast \cdot x}{2p \cdot x} \left[ p_\alpha p_\mu g_{\beta\nu}^\perp - p_\beta p_\mu g_{\alpha\nu}^\perp - p_\alpha p_\nu g_{\beta\mu}^\perp + p_\beta p_\nu g_{\alpha\mu}^\perp \right] \mathcal{T}(\bar{v},px) \non
&+& f_V^\perp m_V^2 \left[ p_\alpha \eta^\ast_{\perp\mu} g_{\beta\nu}^\perp - p_\beta \eta^\ast_{\perp\mu} g_{\alpha\nu}^\perp - p_\alpha \eta^\ast_{\perp\nu} g_{\beta\mu}^\perp + p_\beta \eta^\ast_{\perp\nu} g_{\alpha\mu}^\perp \right] T_1^{(4)}(\bar{v},px) \non
&+& f_V^\perp m_V^2 \left[ p_\mu \eta^\ast_{\perp\alpha} g_{\beta\nu}^\perp - p_\mu \eta^\ast_{\perp\beta} g_{\alpha\nu}^\perp - p_\nu \eta^\ast_{\perp\alpha} g_{\beta\mu}^\perp + p_\nu \eta^\ast_{\perp\beta} g_{\alpha\mu}^\perp \right] T_2^{(4)}(\bar{v},px) \non
&+& \frac{f_V^\perp m_V^2}{p \cdot x}
\left[ p_\alpha p_\mu \eta^\ast_{\perp\beta} x_\nu - p_\beta p_\mu \eta^\ast_{\perp\alpha} x_\nu - p_\alpha p_\nu \eta^\ast_{\perp\beta} x_\mu + p_\beta p_\nu \eta^\ast_{\perp\alpha} x_\mu \right] 
T_3^{(4)}(\bar{v},px) \non
&+& \frac{f_V^\perp m_V^2}{p \cdot x}
\left[ p_\alpha p_\mu \eta^\ast_{\perp\nu} x_\beta - p_\beta p_\mu \eta^\ast_{\perp\nu} x_\alpha - p_\alpha p_\nu \eta^\ast_{\perp\mu} x_\beta + p_\beta p_\nu \eta^\ast_{\perp\mu} x_\alpha \right] T_4^{(4)}(\bar{v},px),
\label{eq:LCDAS-definition4} \\
&~&\langle V(p,\eta^\ast) \vert \bar{q}_1(0) g \tilde{G}_{\mu\nu}(\bar{v}x) \gamma_\alpha \gamma_5 q_2(x) \vert 0\rangle \non
&=& f_V^\parallel m_V \Big[ 
- p_\alpha (p_\mu \eta^\ast_{\perp\nu} - p_\nu \eta^\ast_{\perp\mu}) \mathcal{A}(\bar{v},px) 
+ m_V^2 \frac{\eta^\ast \cdot x}{p \cdot x}(p_\mu g^\perp_{\alpha\nu} - p_\nu g^\perp_{\alpha\mu}) \Phi(\bar{v},px) \non
&+& m_V^2 \frac{\eta^\ast \cdot x}{(p \cdot x)^2} p_\alpha (p_\mu x_\nu - p_\nu x_\mu) \Psi(\bar{v},px) \Big], \label{eq:LCDAS-definition7}\\
&~&\langle V(p,\eta^\ast) \vert \bar{q_1}(0)g G_{\mu\nu}(\bar{v}x)\gamma_\alpha q_2(x) \vert 0\rangle \non
&=& i f_V^\parallel m_V \Big[ 
- p_\alpha (p_\mu \eta^\ast_{\perp\nu} - p_\nu \eta^\ast_{\perp\mu}) \mathcal{V}(\bar{v},px) 
+ m_V^2 \frac{\eta^\ast \cdot x}{p \cdot x}(p_\mu g^\perp_{\alpha\nu} - p_\nu g^\perp_{\alpha\mu}) \tilde{\Phi}(\bar{v},px) \non
&+& m_V^2 \frac{\eta^\ast \cdot x}{(p \cdot x)^2} p_\alpha (p_\mu x_\nu - p_\nu x_\mu) \tilde{\Psi}(\bar{v},px) \Big], \label{eq:LCDAS-definition8}\\
&~&\langle V(p,\eta^\ast) \vert \bar{q_1}(0)g G_{\mu\nu}(\bar{v}x)q_2(x) \vert 0\rangle 
= -i f_V^\perp m_V^2 (\eta^\ast_{\perp\mu} p_\nu - \eta^\ast_{\perp\nu} p_\mu) S(\bar{v},px),
\label{eq:LCDAS-definition5}\\
&~&\langle V(p,\eta^\ast) \vert \bar{s}(x)ig \tilde{G}_{\mu\nu}(\bar{v}x)\gamma_5 q_2(x) \vert 0\rangle 
= - i f_V^\parallel m_V^2 (\eta^\ast_{\perp\mu} p_\nu - \eta^\ast_{\perp\nu} p_\mu) \tilde{S}(\bar{v},px),
\label{eq:LCDAS-definition6}
\eeq
where the auxiliary functions are introduced as 
\beq \mathcal{T}(\bar{v},px)
=\int\mathcal{D}\underline{\alpha}e^{ipx(\alpha_1+\bar{v}\alpha_3)}\mathcal{T}(\alpha_i)\ . \eeq 
The twist-3 and twist-4 DAs read
\beq && \mathcal{V}(\alpha_1,\alpha_3)
=540 \, \zeta_3^\parallel\omega_3^V\left(\alpha_1-(1-\alpha_1-\alpha_3)\right)\alpha_1(1-\alpha_1-\alpha_3)\alpha_3^2,\non
&& \mathcal{A}(\alpha_1,\alpha_3)
=360 \, \zeta_3^\parallel\alpha_1(1-\alpha_1-\alpha_3)\alpha_3^2[1+\omega_3^A\frac{1}{2}(7\alpha_3-3)],\non
&& \mathcal{T}(\alpha_1,\alpha_3)
=540 \, \zeta_3^\parallel\omega_3^T\left(\alpha_1-(1-\alpha_1-\alpha_3)\right)\alpha_1(1-\alpha_1-\alpha_3)\alpha_3^2,
\label{eq:LCDAs3pt3} \eeq
and 
\beq && S(\alpha_1,\alpha_3)
= 30\zeta_4^T(1-\alpha_3)\alpha_3^2,\qquad 
\tilde{S}(\alpha_1,\alpha_3)
= 30 \, \tilde{\zeta}_4^T(1-\alpha_3)\alpha_3^2,\non
&& T_1^{(4)}(\alpha_1,\alpha_3)=
T_3^{(4)}(\alpha_1,\alpha_3)=0,\qquad
T_2^{(4)}(\alpha_1,\alpha_3)
=30 \, \tilde{\zeta}_4^T
\left[ \alpha_1-(1-\alpha_1-\alpha_3) \right]\alpha_3^2,\non
&& T_4^{(4)}(\alpha_1,\alpha_3)
=-30 \, \zeta_4^T
\left[ \alpha_1-(1-\alpha_1-\alpha_3) \right]\alpha_3^2,\non
&& \tilde{\Phi}(\alpha_1,\alpha_3)
= 10 \, (-\zeta_3^\parallel+\zeta_4 )(1-\alpha_3) \alpha_3^2,\qquad 
\tilde{\Psi}(\alpha_1,\alpha_3) 
= 40 \, (2 \zeta_3^\parallel+\zeta_4) \alpha_1(1-\alpha_1-\alpha_3)\alpha_3,\non
&& \Phi(\alpha_1,\alpha_3)
=10 \, (-\zeta_3^\parallel+\zeta_4) \left[(1-\alpha_1-\alpha_3)-\alpha_1 \right] \alpha_3^2,\qquad 
\Psi(\alpha_1,\alpha_3) =0.
\label{eq:LCDAs3pt4}
\eeq

\end{document}